\documentclass[preprint]{aastex}
\shorttitle{MMT observations of new emission galaxies in the SDSS}
\shortauthors{Y. I. Izotov \& T. X. Thuan}

\begin{document}
\title{MMT observations of new
extremely metal-poor emission-line galaxies in the Sloan Digital
Sky Survey}
\author{Yuri I. Izotov}
\affil{Main Astronomical Observatory, National Academy of Sciences of Ukraine,
03680, Kyiv, Ukraine}
\email{izotov@mao.kiev.ua}

\author{Trinh X. Thuan}
\affil{Astronomy Department, University of Virginia,
    Charlottesville, VA 22903}
\email{txt@virginia.edu}

\begin{abstract}
We present 6.5-meter MMT spectrophotometry\footnote{The MMT is operated by 
the MMT Observatory (MMTO), a joint venture of the Smithsonian Institution  
and the University of Arizona.}
of 20 H {\sc ii} regions in 13 extremely metal-poor emission-line 
galaxies selected from the Data Release 5 of the Sloan Digital
Sky Survey to have [O {\sc iii}]$\lambda$4959/H$\beta$ $\la$ 1 and 
[N {\sc ii}]$\lambda$6583/H$\beta$ $\la$ 0.05. 
The electron temperature-sensitive emission line
[O {\sc iii}] $\lambda$4363 is detected in 13 H {\sc ii} regions
allowing a direct abundance determination. The oxygen
abundance in the remaining H {\sc ii} regions is derived using 
a semi-empirical method.
The oxygen abundance of the galaxies in our sample ranges
from 12 + log O/H $\sim$ 7.1 to $\sim$ 7.8, with 10 H {\sc ii}
regions having an oxygen abundance lower than 7.5.
The lowest oxygen abundances, 
12 +log O/H = 7.14 $\pm$ 0.03 and 7.13 $\pm$ 0.07,
are found in two 
H {\sc ii} regions of the blue compact dwarf galaxy 
SDSSJ0956+2849 $\equiv$ DDO 68, making it the
second most-metal deficient emission-line galaxy known, after 
SBS 0335--052W.
\end{abstract}

\section {Introduction}

Extremely metal-deficient emission-line galaxies at low redshifts are the 
most promising young galaxy candidates in the local Universe
\citep{G03,IT04b}. They are important to identify because of 
several reasons. First,  
 studies of their nearly pristine interstellar 
medium (ISM) can shed light on the properties of the primordial ISM at 
the time of galaxy formation.     
 It appears now that even the most metal-deficient galaxies in the local 
Universe formed from matter which was already pre-enriched by a previous star 
formation episode, e.g. by Population III stars
\citep{T05}. It is thus quite important to establish firmly 
the level of this pre-enrichment
by searching for the most metal-deficient emission-line galaxies. 
Second, because they have not undergone much chemical evolution, 
these galaxies
are also the best objects for the determination of the primordial He 
abundance and for constraining cosmological models 
\citep[e.g. ][]{IT04a,ITS07}. 
Third, in the hierarchical picture of galaxy 
formation, large galaxies form from the assembly of small dwarf 
galaxies.  
While much progress has been made in finding large populations of galaxies at 
high redshifts \citep[$z\ga\,3$, ][]{steidel03}, 
truly young galaxies in the process of forming remain 
elusive in the distant universe. 
The spectra of those far-away galaxies generally indicate the presence of a 
substantial amount of heavy elements, 
indicating previous star formation and metal enrichment. 
Therefore, extremely metal-deficient dwarf galaxies  
 are possibly the closest examples we can find of the elementary 
primordial units from which galaxies formed. 
Their relative proximity allows studies of their 
stellar, gas and dust content with a sensitivity, 
spectral and spatial resolution that faint distant high-redshift galaxies 
do not permit.

Extremely metal-deficient emission-line galaxies are however very rare. 
Many surveys have been carried out 
to search for such galaxies without significant success. For more than three
decades,
one of the first blue compact dwarf (BCD) galaxies discovered, I Zw 18 
\citep{SS70} continued to hold the record as the most metal-deficient 
emission-line galaxy known, with
an oxygen abundance 12 + log O/H = 7.17 $\pm$ 0.01 in its 
northwestern component 
and 7.22 $\pm$ 0.02 in its southeastern component \citep{TI05}. 
Only very recently, has I Zw 18 been displaced by the BCD 
SBS 0335--052W. This galaxy, with an oxygen abundance 
12 + log O/H = 7.12 $\pm$ 0.03, is now the emission-line galaxy 
with the lowest metallicity known \citep{I05}. 

Because of the scarcity of extremely low-metallicity galaxies such as I Zw 18 
and SBS 0335--052W, we stand a better
chance of finding them in very large spectroscopic surveys. 
One of the best surveys suitable for
such a search is the Sloan Digital Sky Survey (SDSS) \citep{Y00}. 
However, despite
intensive studies of galaxies with a detected 
temperature-sensitive [O {\sc iii}]
$\lambda$4363 emission line in their spectra, no emission-line galaxy
with an oxygen abundance as low as that of I Zw 18 has been 
discovered in the SDSS Data Release 3 (DR3) and earlier releases. 
The lowest-metallicity emission-line galaxies found so far in these releases 
have oxygen abundances 12 + log O/H $>$ 7.4 \citep{K04a,K04b,I04,I06}. 
Only recently, \citet{I06b} have shown that two galaxies, J2104$-$0035 and
J0113$+$0052, selected from the SDSS Data Release 4 (DR4), are very
metal-poor, with 12+logO/H = 7.26 $\pm$ 0.03 and 7.17 $\pm$ 0.09, 
respectively.

In order to find new candidates for extremely metal-deficient 
emission-line galaxies, we have carried out a systematic search 
for such objects in the SDSS Data Release 5 (DR5) \citep{A07}.
We have chosen them 
on the basis of the relative fluxes of selected  
emission lines, as in \citet{I06b}. 
All known extremely metal-deficient emission-line galaxies 
are characterized by relatively weak (compared to 
H$\beta$) [O {\sc ii}] $\lambda$3727, [O {\sc iii}] $\lambda$4959, 
$\lambda$5007 and [N {\sc ii}] $\lambda$6583 emission 
lines \citep[e.g. ][]{IT98a,IT98b,I05,P05,I06b}.
These spectral properties select out uniquely low-metallicity 
dwarfs since no other type of galaxy possesses them. 
In contrast to previous studies \citep{K04a,K04b,I04,I06} which focus 
exclusively 
on objects with a detected [O {\sc iii}] $\lambda$4363 emission line,
we have also considered objects which satisfy the criteria described above, 
but with spectra  
where [O {\sc iii}] $\lambda$4363 is weak or not detected.
Since the [O {\sc ii}] $\lambda$3727 line is out of the observed wavelength 
range in SDSS spectra of galaxies with a 
redshift $z$ lower than 0.02, we use two criteria,  
[O {\sc iii}]$\lambda$4959/H$\beta$ $\la$ 1 and 
[N {\sc ii}]$\lambda$6583/H$\beta$ $\la$ 0.05,
 to pick out  $\sim$ 100 galaxies from the DR5.
The efficiency of this technique to pick out 
extremely low-metallicity galaxy candidates has been demonstrated 
by \citet{I06b}. 

While the SDSS spectra allow us to select very low 
metallicity galaxies, we need additional spectral 
observations for the following reasons: 
1) a spectrum that goes further into the blue 
wavelength range is required to detect the  [O {\sc ii}] $\lambda$3727 line. 
For a precise oxygen abundance determination, 
this line is needed  
to provide information on the singly ionized ionic population of oxygen. 
In principle, the [O {\sc ii}]
$\lambda$7320, 7330 emission lines can be used instead. However,
these lines are very weak or not detected in the SDSS spectra of 
low-metallicity candidates. 2) A better signal-to-noise ratio
spectrum may allow the detection of 
 a weak [O {\sc iii}] $\lambda$4363 emission 
line, which would permit a direct determination of the electron temperature.
3) Very often, low-metallicity candidates possess two or more 
H {\sc ii} regions with different degrees of excitation. However, 
SDSS spectra are usually obtained for only one H {\sc ii} region,
chosen sometimes not in the most optimal way. A case in point is that of 
the galaxy J2104$-$0035. It
 has two H {\sc ii} regions.
Only the spectrum of the lower excitation one, with 
no [O {\sc iii}] $\lambda$4363 emission, is 
available in the SDSS data base. \citet{I06b} obtained a 
ESO 3.6 m spectrum of 
the second H {\sc ii} region which turned out to have a higher 
excitation and a clearly
detected [O {\sc iii}] $\lambda$4363 emission line, allowing a reliable 
abundance determination. This new observation of J2104$-$0035 
has shown the galaxy to be very metal-poor.

For these reasons, we have started an observational program using the  
MMT to obtain new spectroscopic observations of a subsample of 
SDSS metal-poor galaxy candidates. The observations and
data reduction are discussed in \S\ref{S2}. The element abundances are
derived in \S\ref{S3}. Our main findings are given in
\S\ref{S4}.

\section {Observations and Data Reduction \label{S2}}

\begin{figure}
\figurenum{1}
\epsscale{0.7}
\plotone{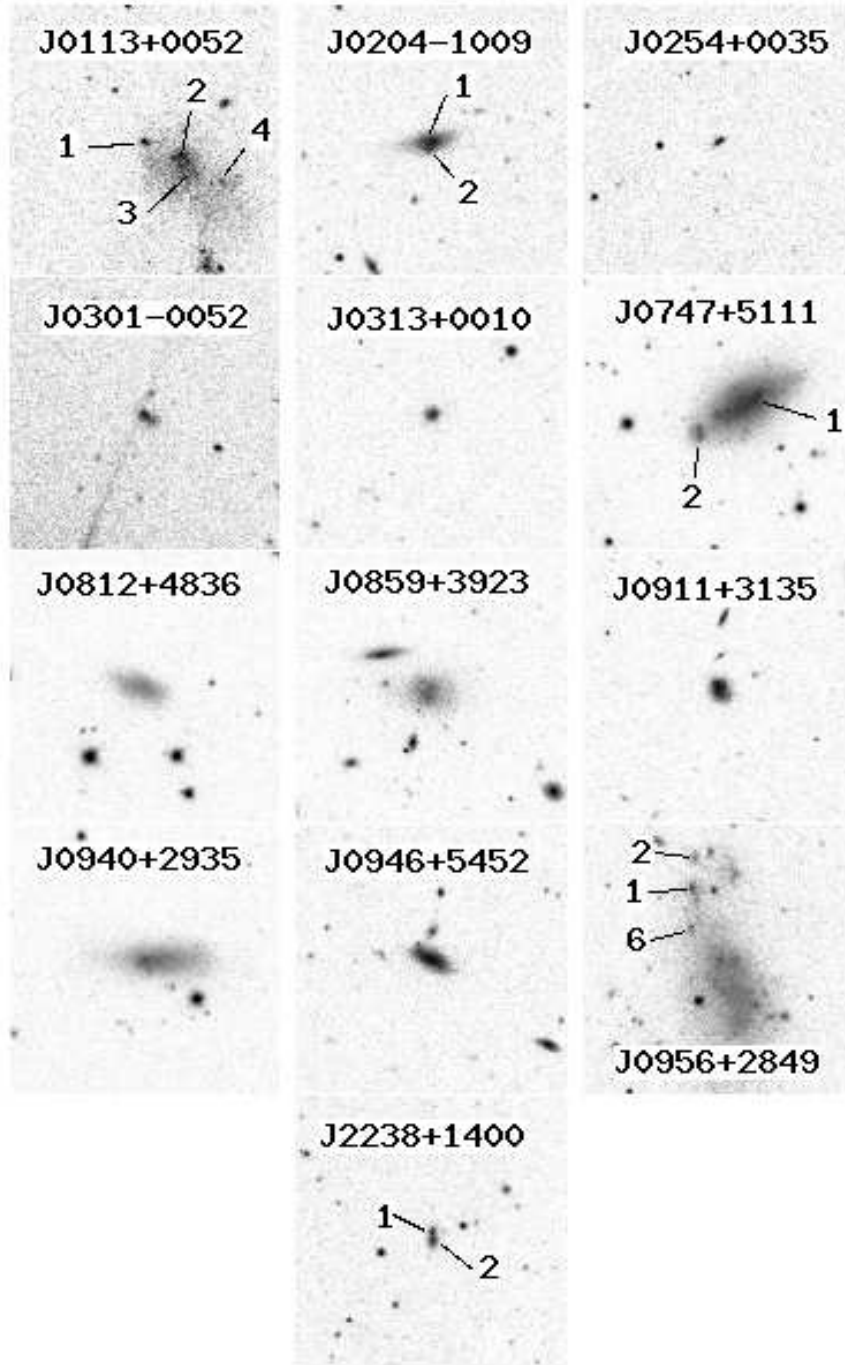}
\caption{SDSS images of galaxies in MMT subsample.}
\label{fig1}
\end{figure}

We have obtained new high signal-to-noise ratio spectrophotometric 
observations of 20 H {\sc ii} regions in 13 emission-line galaxies with
the 6.5-meter MMT on the nights of 2006 December 15 -- 16.
The galaxies are listed in Table \ref{tab1} in order of increasing
right ascension, along with some of their general properties such as 
coordinates, redshifts, identifications of their
SDSS spectra, and other designations. The SDSS images of the observed 
galaxies are shown in Fig. \ref{fig1}. Labels of individual  
H {\sc ii} regions are shown when 
several of them have been observed within the same galaxy. 
We show in Fig. \ref{fig2} (available only in the on-line version) 
the SDSS spectra  
used to select the galaxies from the DR5. It is seen from
the figure that the [O {\sc ii}] $\lambda$3727 emission line is out of the  
observed spectral range and that 
the [N {\sc ii}] $\lambda$6584 emission line is weak in 
all spectra. 

Ten out of the thirteen galaxies listed in 
Table \ref{tab1} have been chosen from the DR5 with the selection criteria
described before. Their oxygen abundances are not known before this work.
 The 3 remaining galaxies satisfy also 
the selection criteria, except for J2238+1400 where 
[O {\sc iii}]$\lambda$4959/H$\beta$ $\sim$ 1.6. 
However, their oxygen abundances 
have been determined before. Two galaxies have been included in our observing 
program because 
they are among the five most metal-deficient emission-line galaxies known,
and we believe that more information can be obtained about them with 
supplementary observations.     
\citet{I06b} have obtained 12+log(O/H) = 7.17$\pm$0.09 for 
J0113+0052$\equiv$UGC 772, and \citet{P05} have measured  
12+log(O/H) = 7.21$\pm$0.07 for J0946+5452$\equiv$DDO 68.
 We wish to improve the
abundance determination in these two galaxies with higher
signal-to-noise ratio observations. The 
third galaxy is J2238+1400$\equiv$HS 2236+1344 which 
has been studied spectroscopically by \citet{U03}, \citet{IT04a} and
\citet{G07}. \citet{IT04a} have found in its spectrum a strong high-ionization
[Fe {\sc v}] $\lambda$4227 emission line, with an ionization 
potential of 4 Ryd. We have included this galaxy in our observing 
program to search for the high-ionization
[Ne {\sc v}] $\lambda$3346, 3425 line emission which has an ionization 
potential of $\sim$ 7 Ryd, and to check for the consistency of the 
abundance determinations for this object by different authors,
using different telescopes. 

With the exception of J0254+0035 and J2238+1400,
all galaxies in Table \ref{tab1} 
are at a distance $\la$ 30 Mpc. Five of them 
are within a distance of 10 Mpc, which makes them ideal for a 
study of their resolved stellar populations with 
the {\sl Hubble Space Telescope}.

   All observations have been made with the Blue Channel of the MMT 
spectrograph.  The log of the observations is given in Table \ref{tab2}.
We used a 1\farcs5$\times$300$''$ slit and a 800 grooves/mm grating in first
order. The above instrumental set-up gave a spatial scale
along the slit of 0\farcs6 pixel$^{-1}$, a scale perpendicular to the slit
of 0.75\AA\ pixel$^{-1}$, a spectral range of 3200--5200\AA\ and a spectral
resolution of $\sim$ 3\AA\ (FWHM). 
The seeing was in the range 1$\arcsec$--1\farcs5. 
Total exposure times varied 
between 30 and 45 minutes. Each exposure was broken up into 2--3 
subexposures, not exceeding 15 minutes, to allow for removal of cosmic rays. 
Several objects were observed at low airmasses $<$ 1.3 or with the slit
oriented along the parallactic angle. The latter observations are labeled (P)
in Table \ref{tab2}. The effect of atmospheric refraction for these 
observations is small. However, it can be important in the spectra of
galaxies observed at high airmasses $>$ 1.3. Fortunately, strong hydrogen
lines are observed in some spectra (i.e. those of J0204--1009, J0747+5111) 
and correction for interstellar extinction
with the use of those lines will automatically take into account the effect
of atmospheric refraction. This is because interstellar extinction 
correction is performed in such a way so that the intensities of all observed
hydrogen lines after correction are as close as possible to their
theoretical recombination values, for a given electron temperature. However, 
in some other spectra obtained at high airmasses 
(those of J0301--0052, J0812+4836,
J0911+3135, J0940+2935 and J0946+5452), the hydrogen lines are weaker, 
making such correction less certain.
     Three Kitt Peak IRS spectroscopic standard stars, G191B2B, Feige 110 and 
BD +28 4211 have been observed
for flux calibration. Spectra of He--Ne--Ar comparison arcs were obtained
before and after each observation to calibrate the wavelength scale.

    The two-dimensional spectra were bias subtracted and flat-field corrected
using IRAF\footnote{IRAF is distributed by National Optical Astronomical 
Observatory, which is operated by the Association of Universities for 
Research in Astronomy, Inc., under cooperative agreement with the National 
Science Foundation.}. We then use the IRAF
software routines IDENTIFY, REIDENTIFY, FITCOORD, TRANSFORM to 
perform wavelength
calibration and correct for distortion and tilt for each frame. 
One-dimensional spectra were then extracted from each frame using the APALL 
routine. Before extraction, distinct two-dimensional 
spectra of the same H {\sc ii} region
were carefully aligned using the spatial locations of the brightest part in
each spectrum, so that spectra were extracted at the same positions in all
subexposures. For all objects, we extracted the 
brightest part of the BCD, corresponding to a different spatial size
for each object.  In all cases, 6\arcsec$\times$1\farcs5 extraction apertures
were used.
All extracted spectra from the same object were then co-added. 
We have summed the individual spectra 
from each subexposure after removal of the cosmic rays hits with the IRAF
routine CRMEDIAN. The spectra obtained from each subexposure
were also checked for cosmic rays hits at the location of strong 
emission lines, but none was found.

The sensitivity curve was obtained by 
fitting with a high-order polynomial the observed spectral energy 
distribution of the bright hot white dwarf standard stars G191B2B, Feige 110 
and BD +28 4211. Because the spectra of these stars have only a small number of
a relatively weak absorption features, their spectral energy distributions are 
known with a very good accuracy \citep{O90}. 
Moreover, the response function of the CCD detector is smooth, so we could
derive a sensitivity curve with a precision better than 1\% over the
whole optical range. 

The spectra for the 20 H {\sc ii} regions observed with
the MMT are shown in Figure \ref{fig3}. These spectra 
have been reduced to zero redshift and corrected for extinction. 

   The observed line fluxes $F$($\lambda$), normalized to $F$(H$\beta$) 
and multiplied by a factor of 100, and
their errors, for the 20 H {\sc ii} regions shown in Fig. \ref{fig3} are 
given in Table \ref{tab3}, available only in the electronic version on line. 
They were measured using the IRAF SPLOT routine.
The line flux errors listed include statistical errors derived with
SPLOT from non-flux calibrated spectra, in addition to errors introduced
in the standard star absolute flux calibration, which we set to 1\% of the
line fluxes. These errors will be later propagated into the calculation
of abundance errors.
The line fluxes were corrected for both reddening (using 
the extinction curve of \citet {W58}) 
and underlying hydrogen stellar absorption derived simultaneously by an 
iterative procedure as described in \citet{ITL94}. 
The extinction coefficient is defined as $C$(H$\beta$) = 1.47$E(B-V)$,
where $E(B-V)$ = $A(V)$/3.2 and $A(V)$ is the extinction in the $V$ band
\citep{A84}.
The corrected line 
fluxes 100$\times$$I$($\lambda$)/$I$(H$\beta$), equivalent widths 
EW($\lambda$), extinction
coefficients $C$(H$\beta$), and equivalent widths EW(abs) of the hydrogen
absorption stellar lines are also given in Table \ref{tab3}, along with the 
uncorrected H$\beta$ fluxes.

\section {Physical Conditions and Element Abundances \label{S3}}

   To determine element abundances, we follow generally 
the procedures of \citet{ITL94,ITL97} and \citet{TIL95}.
We adopt a two-zone photoionized H {\sc ii}
region model: a high-ionization zone with temperature $T_e$(O {\sc iii}), 
where [O {\sc iii}] and [Ne {\sc iii}] lines originate, and a 
low-ionization zone with temperature $T_e$(O {\sc ii}), where [O {\sc ii}], 
[N {\sc ii}], [S {\sc ii}] and [Fe {\sc iii}] lines originate. 
In the H {\sc ii} regions with a detected [O {\sc iii}] $\lambda$4363
emission line, the temperature $T_e$(O {\sc iii}) is calculated using the 
``direct'' method based on the 
[O {\sc iii}] $\lambda$4363/($\lambda$4959+$\lambda$5007) line ratio.
In H {\sc ii} regions where the [O {\sc iii}] $\lambda$4363 emission line 
is not detected,
we used an empirical relation obtained from the
photoionized models of \citet{SI03} and based on the 
more readily observable [O {\sc ii}] $\lambda$3727 and 
[O {\sc iii}] $\lambda\lambda$ 4959, 5007 lines :
\begin{equation}
t_e({\rm O}\ {\rm III})=
-1.36854\log\left[\frac{I(\lambda3727)+I(\lambda4959)+I(\lambda5007)}{I({\rm H}\beta)}\right]+2.62577, 
\label{eq:toiii}
\end{equation}
where $t_e$(O {\sc iii}) = 10$^{-4}$$T_e$(O {\sc iii}). We adopt 
the error in the temperature determined by 
Eq. \ref{eq:toiii} to be equal to the dispersion
of $t_e$s of individual H {\sc ii} region models about the least-square fit.
This dispersion is $\sim$ 1000K. The error is then 
propagated into the calculation
of abundance errors. We will refer hereafter to this method 
as the ``semi-empirical method'', to emphasize the fact that only the 
electron temperature is derived empirically, while 
abundances are calculated in the same way as those of H {\sc ii} regions
with a detected [O {\sc iii}] $\lambda$4363 emission line, 
and to distinguish it from the ``empirical'' methods 
where the oxygen abundance is derived directly from the strong oxygen line 
intensities \citep[e.g., ][]{PT05}. 

For $T_e$(O {\sc ii}), we use
the relation between the electron temperatures $T_e$(O {\sc iii}) and
$T_e$(O {\sc ii}) obtained by \citet{I06} from
the H {\sc ii} photoionization models of \citet{SI03}. These are based on
more recent stellar atmosphere models and improved
atomic data as compared to the \citet{S90} models. 
As the [S {\sc ii}] $\lambda$6717 and $\lambda$6731 emission 
lines are not in the observed wavelength region of the MMT spectra, 
$N_e$(S {\sc ii}) was set to 10 cm$^{-3}$ for all H {\sc ii} regions. 
This is justified as element abundances do not depend on $N_e$ as long as
its value is lower than $\sim$ 10$^4$ -- 10$^5$ cm$^{-1}$, which is the 
case for the vast majority of  
of the H {\sc ii} regions in the emission-line galaxies
considered here 
\citep[e.g. ][]{I06}. Ionic 
and total heavy element abundances are derived 
using expressions for ionic abundances and ionization correction 
factors obtained by \citet{I06}.
The element abundances are given in 
Table \ref{tab4} (available only in the electronic version on line)
 along with the adopted electron temperatures for
different ions.

Consider first the 
results obtained with the direct method 
for the 13 H {\sc ii} regions with a detected
[O {\sc iii}] $\lambda$4363 emission line. The derived 
oxygen abundances for these
H {\sc ii} regions are shown in the second column of Table \ref{tab5}.
The galaxy J2238+1400, with two high-excitation H {\sc ii} regions,
was observed to check for the presence of 
the high-ionization [Ne {\sc v}] $\lambda$3346,
$\lambda$3425 emission lines in its spectra. No such lines were detected
(Fig. \ref{fig3}t,u). Thus, despite the presence of the [Fe {\sc v}]
$\lambda$4227 emission line in the spectrum of its H {\sc ii} region
No. 1 (Table \ref{tab3}), implying ionizing radiation with photon
energies $>$ 4 Ryd, no ionizing radiation with photon 
energies above 7 Ryd is present in this galaxy. This is in contrast to the
situation in the three low-metallicity BCDs SBS 0335--052E, Tol 1214--277 and
HS 0837+4717 known thus far to contain both 
[Fe {\sc v}] and [Ne {\sc v}] emission
\citep{TI05}. The oxygen abundances 12 + log O/H = 7.45 and 7.56 
derived respectively in regions 
J2238+1400 No.1 and J2238+1400 No.2 are in good agreement with
the values 7.49 and 7.58 derived by \citet{G07} for the same H {\sc ii}
regions. A very similar oxygen abundance of 7.47 was obtained for region No.1
by \citet{U03} and \citet{IT04a}. This shows that the abundances obtained 
for this galaxy by different authors, using different telescopes, are very 
consistent.
Two other galaxies, J0113+0052 and J0956+2849, 
have been observed before.
\citet{I06b} have obtained 12 + log O/H = 7.17 $\pm$ 0.09 for J0113+0052 No.1.
With our higher signal-to-noise ratio spectrum, we derive 12 + log O/H = 
7.24 $\pm$ 0.05. \citet{P05} have obtained 12 + log O/H = 7.23 $\pm$ 0.06
and 7.21 $\pm$ 0.07 for H {\sc ii} regions No.1 and No.2 in
J0956+2849 \citep[we follow ][ for the nomenclature of the 
H {\sc ii} regions]{P05}. We derive
12 + log O/H = 7.14 $\pm$ 0.03 and 7.13 $\pm$ 0.07 from our higher
resolution and higher signal-to-noise ratio spectra of the same H {\sc ii}
regions. The new measured abundances 
make J0956+2849 the second most metal-deficient 
emission-line galaxy known after SBS 0335--052W \citep{I05}. It is 
more metal-deficient than I Zw 18 NW (12 + log O/H = 7.17 $\pm$ 0.01),
I Zw 18 SE (12 + log O/H = 7.22 $\pm$ 0.01) and SBS 0335--052E 
(12 + log O/H = 7.31 $\pm$ 0.01) \citep{TI05}.
Other galaxies
with a detected [O {\sc iii}] $\lambda$4363 have not been observed
previously. They have higher oxygen abundances, in the range 
12 + log O/H = 7.51 -- 7.75.

While the determination of oxygen abundance in H {\sc ii} regions 
with a detected
[O {\sc iii}] $\lambda$4363 via the direct method 
is straightforward, 
the determination of oxygen abundances in those  H {\sc ii} regions 
with no detected
[O {\sc iii}] $\lambda$4363 emission has to rely on semi-empirical or 
empirical methods based on the intensities of the 
strong [O {\sc ii}] $\lambda$3727
and [O {\sc iii}] $\lambda$4959, $\lambda$5007 nebular emission lines.
In order to evaluate the accuracy of the abundances based on 
these semi-empirical or empirical methods, 
we have compared oxygen abundances derived with these alternative methods with
those obtained by the direct method for the 
H {\sc ii} regions with detected [O {\sc iii}]
$\lambda$4363 emission in our sample.

First, we compare the oxygen abundances derived with the direct method 
(column 2 in Table \ref{tab5}) with those obtained by 
the semi-empirical method described above (column 6 in Table \ref{tab5}).
We find a general good agreement between the 
two methods. Exceptions are the two 
high excitation H {\sc ii} regions
in J2238+1400. In these cases, the electron temperatures 
derived by the semi-empirical method (Eq. \ref{eq:toiii}) 
are $\sim$ 4000 -- 5000K lower than those derived from the
[O {\sc iii}] $\lambda$4363/($\lambda$4959+$\lambda$5007) ratio, yielding  
significantly higher oxygen abundances. 
These discrepancies may indicate that 
these two H {\sc ii} regions have anomalous
properties. This may be the case if 
[O {\sc iii}] $\lambda$4363 emission is enhanced
by some mechanism that is different from stellar radiation heating
such as shock heating, or if the [O {\sc iii}] 
$\lambda$4363/($\lambda$4959+$\lambda$5007) ratio is 
artificially enhanced by a reduction of the 
fluxes of the [O {\sc iii}] $\lambda$4959 and $\lambda$5007
emission lines by collisional de-excitation in a dense 
interstellar medium ($N_e$ $\ga$ 10$^5$ cm$^{-3}$), as discussed by 
\citet{T96} in the case of   
the BCD Mrk 996. In these conditions, the oxygen 
abundance derived by the direct method will be lower than  the true value.
 Another possible mechanism for the 
enhancement of [O {\sc iii}] $\lambda$4363 emission is 
nonthermal radiation from a hidden active galactic nucleus (AGN). 
 However, we do not favor this mechanism because 
the optical spectra of both H {\sc ii} regions in J2238+1400 
do not show the usual features that are characteristic of an  
AGN Seyfert 2 optical spectrum: the [Ne {\sc v}] 
$\lambda$3346, 3425 emission lines are not detected (see the inset in 
Fig. \ref{fig3}t), 
the He {\sc ii} $\lambda$4686 emission line is weak
(its intensity is only $\sim$ 1\% of that of H$\beta$,
Table \ref{tab3}), and other lines, such as [O {\sc i}] $\lambda$6300,
[N {\sc ii}]$\lambda$6584 and [S {\sc ii}] $\lambda$6717, 6731 are many times
weaker than those in a typical Seyfert 2 galaxy (see Fig. \ref{fig2}m).
We note however that Spitzer observations of several BCDs 
\citep{Hunt06,T07} have shown 
that the [O {\sc iv}] $\lambda$25.9 \micron\ emission line, 
with an ionization 
potential of $\sim$ 4 Ryd is present in the MIR spectrum, while 
the [Fe {\sc iv}] $\lambda$4227 and He {\sc ii} $\lambda$4686 
emission lines with about the same ionization potential are 
conspicuously absent from the optical spectrum. 
This suggests that, in those BCDs, there may be a dust-enshrouded AGN that 
is optically invisible and which 
may be responsible for the hard radiation giving rise to the MIR line.
If we exclude the two H {\sc ii} regions in J2238+1400, 
the average difference 
between the semi-empirical abundances and those derived by the direct method
is 0.06 dex. 
The semi-empirical method also 
gives consistent oxygen abundances for multiple H {\sc ii} regions
within the same galaxy, including those where the [O {\sc iii}]
$\lambda$4363 emission line is not detected (see the galaxies
J0113+0052 and in J0956+2849). 
 
Next, we compare the abundances determined by the direct method with those 
derived using empirical methods. 
Recently, several groups have proposed empirical relations 
for the determination
of 12 + log O/H using the nebular [O {\sc ii}] $\lambda$3727 and [O {\sc iii}]
$\lambda$4959, $\lambda$5007 emission line intensities:
\begin{equation}
12+\log\frac{\rm O}{\rm H}=
\frac{R_3+106.4P+106.8P^2-3.40P^3}{17.72P+6.60P^2+6.95P^3-0.302R_3},
\label{eq:o1}
\end{equation}
\citep{PT05},
\begin{equation}
\log R_{23}=
1.2299-4.1926y+1.0246y^2-0.063169y^3,
\label{eq:o2}
\end{equation}
\citep{N06}, and
\begin{equation}
12+\log\frac{\rm O}{\rm H}=
6.486+1.401\log R_{23},
\label{eq:o3}
\end{equation}
\citep{Y07}. 

In Eqs. \ref{eq:o1} -- \ref{eq:o3},
$R_3$ = [$I$($\lambda$4959) + $I$($\lambda$5007)]/$I$(H$\beta$),
$R_{32}$ = $R_3$ + $I$($\lambda$3727)/$I$(H$\beta$), $P$ = $R_3$/$R_{32}$,
$y$=12+log O/H.
We have derived oxygen abundances for all H {\sc ii}
regions in our sample using the above equations.
The results are shown in columns 3 -- 5 of Table \ref{tab5}.
Comparison with the abundances derived by the direct method shows that the 
empirical method of \citet{PT05} does not work well in some cases.
In particular, while the direct method gives abundances that are 
consistent and in a narrow range 
for the multiple H {\sc ii} regions 
in J0747+5111 and J0956+2849, the spread of the  
abundances for the same H {\sc ii} regions derived with the 
Pilyugin-Thuan relation is much larger (column 3 of Table \ref{tab5}). 
The latter relation also gives
a larger spread of oxygen abundances for the 
multiple H {\sc ii} regions in
J0113+0052, as compared to those obtained with the semi-empirical method.
The average difference 
between the empirical abundances derived with the Pilyugin-Thuan relation
and those derived by the direct method
is 0.10 dex, excluding the two H {\sc ii} regions in J2238+1400.
Clearly, the empirical relation derived by \citet{PT05} is not as accurate
as the semi-empirical method 
in the extremely low metallicity regime 
considered here. This is because the 
galaxy sample used by \citet{PT05} to calibrate 
their relation contained very few galaxies with 
12 + log O/H $\la$ 7.5. The empirical relation derived by \citet{N06} 
gives consistent abundances in different
H {\sc ii} regions of the same galaxy (column 4 of Table \ref{tab5}). 
However, these abundances are 
systematically lower than those derived by the direct method. The average 
difference between the empirical abundances derived using the \citet{N06} 
relation and those derived 
by the direct method
is 0.13 dex, excluding the two H {\sc ii} regions in J2238+1400. The
empirical relation derived 
by \citet{Y07} appears to give 
abundances that are most consistent with those derived by 
the direct method (column 5 of Table \ref{tab5}).
The average 
difference between the empirical abundances derived with the 
relation by \citet{Y07} and those derived 
by the direct method
is 0.07 dex, only slightly higher than that between the direct 
and semi-empirical methods.


Thus, we conclude that for our low-metallicity objects, 
the semi-empirical method is more appropriate
for oxygen abundance determination because it gives more consistent
abundances as compared to empirical methods. Therefore, for the H {\sc ii}
regions with non-detected [O {\sc iii}] $\lambda$4363 emission, 
we adopt the oxygen abundances derived with the semi-empirical method, and for
the remaining H {\sc ii} regions those derived with 
the direct method. However, it should be kept in mind that the 
oxygen abundances
derived with the direct method 
for the two H {\sc ii} regions of J2238+1400 may be underestimated.

Examination of Table \ref{tab5} shows 
that our sample contains at least 10 H {\sc ii} regions 
with oxygen abundance less than 7.5. However, no H {\sc ii} region with an
oxygen abundance 12 + logO/H $<$ 7.1 has been found. This supports the idea
discussed by, e.g., 
\citet{T05} that the matter from which dwarf emission-line galaxies
formed was pre-enriched to the level 12 + log O/H $\ga$ 7.0 \citep[or $\sim$
2\% of the abundance 12 + log O/H = 8.65 of the Sun, ][]{asplund05}.
Based on FUSE spectroscopic data, 
\citet{T05} showed that  
BCDs spanning a wide range in ionized gas metallicities all have  
H {\sc i} envelopes with about the same 
neutral gas metallicity of $\sim$7.0. This is also the metallicity  
found in Ly$\alpha$ absorbers. Taken together, the available data suggest
 that there may have been 
previous enrichment of the primordial neutral gas to a common 
metallicity level, possibly by Population III stars.

\section{Conclusions \label{S4}}

We present spectroscopic observations with the 6.5m MMT of
a sample of 20 H {\sc ii} regions in 13 dwarf emission-line
galaxies. These galaxies were selected from the Data Release 5 (DR5)
of the Sloan Digital Sky Survey (SDSS) using the two criteria  
[O {\sc iii}]$\lambda$4959/H$\beta$ $\la$ 1 and 
[N {\sc ii}]$\lambda$6583/H$\beta$ $\la$ 0.05.
These spectral properties select out extremely low-metallicity
galaxies,  with oxygen
abundances comparable to those of the most metal-deficient
emission-line galaxies known, SBS 0335--052W and I Zw 18. 
The above criteria select out  $\sim$ 100 galaxies from the DR5, of which 
the 13 objects discussed here form a subsample. 

We find that 10 H {\sc ii} regions have oxygen abundances
12 + log O/H lower than 7.5. We confirm the very low oxygen abundance 
found previously in two galaxies, J0113+0052$\equiv$UGC 772 and 
J0956+2849$\equiv$DDO 68,
by \citet{I06b} and \citet{P05} respectively.
In particular, we find that the oxygen abundance in the brightest
H {\sc ii} region of J0956+2849 is 7.14 $\pm$ 0.03, making
this galaxy the second most-metal deficient emission-line galaxy known,
after SBS 0335--052W, and ahead of I Zw 18 NW, I Zw 18 SE and 
SBS 0335--052 E. 
However, no H {\sc ii} region with an 
oxygen abundance 12 + logO/H $<$ 7.1 has been found. 
The existing data on extremely metal-deficient emission-line galaxies 
appears to suggest the existence of 
an oxygen abundance floor. This supports the idea
that the matter from which dwarf emission-line galaxies
formed was pre-enriched to a level 12 + log O/H $\sim$ 7.0 
\citep[e.g., ][]{T05}.

\acknowledgements

The MMT time was available thanks to a grant from the 
Frank Levinson Fund of the Peninsula Community Foundation 
to the Astronomy Department of the University of Virginia.
The research has been supported by NSF grant AST-02-05785.
Y.I.I. thanks the hospitality of the Astronomy Department of 
the University of Virginia. 
    Funding for the Sloan Digital Sky Survey (SDSS) and SDSS-II has been 
provided by the Alfred P. Sloan Foundation, the Participating Institutions, 
the National Science Foundation, the U.S. Department of Energy, the National 
Aeronautics and Space Administration, the Japanese Monbukagakusho, and the 
Max Planck Society, and the Higher Education Funding Council for England. 
 


\clearpage

\begin{deluxetable}{lccccl}
\tabletypesize{\footnotesize}
\tablenum{1}
\tablecolumns{6}
\tablewidth{0pt}
\tablecaption{General Characteristics of Galaxies \label{tab1}}
\tablehead{
\colhead{SDSS Name}&\colhead{R.A. (J2000.0)}&\colhead{DEC. (J2000.0)}
&\colhead{Redshift}&\colhead{SDSS Spectrum ID}&\colhead{Other Names} 
}
\startdata
SDSSJ0113$+$0052&01 13 40.44&$+$00 52 39.2&0.0037630&53001-1499-525&UGC 772 \\
SDSSJ0204$-$1009&02 04 25.61&$-$10 09 35.0&0.0063795&52149-0666-088&KUG 0201$-$103 \\
SDSSJ0254$+$0035&02 54 28.94&$+$00 35 50.5&0.0148437&53035-1512-400& \\
SDSSJ0301$-$0052&03 01 49.01&$-$00 52 57.4&0.0072615&52616-1067-204& \\
SDSSJ0313$+$0010&03 13 01.60&$+$00 10 40.2&0.0077796&52203-0710-597& \\
SDSSJ0747$+$5111&07 47 33.18&$+$51 11 24.8&0.0014436&53327-1869-282&KUG 0743$+$513 \\
SDSSJ0812$+$4836&08 12 39.53&$+$48 36 45.5&0.0017567&51885-0440-170& \\
SDSSJ0859$+$3923&08 59 46.93&$+$39 23 05.6&0.0019601&52669-1198-590& \\
SDSSJ0911$+$3135&09 11 59.42&$+$31 35 35.9&0.0025028&52976-1591-097& \\
SDSSJ0940$+$2935&09 40 12.84&$+$29 35 30.3&0.0018161&53415-1942-055&KUG 0937$+$298 \\
SDSSJ0946$+$5452&09 46 22.87&$+$54 52 08.4&0.0054097&52282-0769-376&KUG 0942$+$551 \\
SDSSJ0956$+$2849&09 56 46.05&$+$28 49 43.8&0.0016006&53431-1947-040&DDO 68 \\
SDSSJ2238$+$1400&22 38 31.12&$+$14 00 29.8&0.0206160&53239-1893-476&HS 2236+1344 \\
\enddata
\end{deluxetable}

\clearpage

\begin{deluxetable}{lrrcl}
\tabletypesize{\footnotesize}
\tablenum{2}
\tablecolumns{5}
\tablewidth{0pt}
\tablecaption{Journal of Observations \label{tab2}}
\tablehead{
\colhead{SDSS Name}&\colhead{Date}&\colhead{Exposure}
&\colhead{Airmass}&\colhead{P.A.}
}
\startdata
SDSSJ0113$+$0052&2006, 15 Dec&2460&1.17&$+$63.0 \\
SDSSJ0204$-$1009&      15 Dec&1800&1.40&$-$11.3 \\
SDSSJ0254$+$0035&      15 Dec&2700&1.18&$-$22.5 (P) \\
SDSSJ0301$-$0052&      16 Dec&2700&1.41&$+$48.4 \\
SDSSJ0313$+$0010&      15 Dec&1800&1.18&$+$04.8 (P) \\
SDSSJ0747$+$5111&      16 Dec&2700&1.78&$-$59.4 \\
SDSSJ0812$+$4836&      15 Dec&1800&1.38&$+$67.3 \\
SDSSJ0859$+$3923&      15 Dec&1800&1.44&$-$79.0 (P) \\
SDSSJ0911$+$3135&      16 Dec&1800&1.42&$+$00.0 \\
SDSSJ0940$+$2935&      16 Dec&1800&1.39&$+$24.2 \\
SDSSJ0946$+$5452&      16 Dec&1800&1.71&$+$58.2 \\
SDSSJ0956$+$2849&      16 Dec&2700&1.29&$-$00.8 \\
SDSSJ2238$+$1400&      15 Dec&1800&1.17&$+$00.0 \\
\enddata
\end{deluxetable}

\clearpage

  \begin{deluxetable}{lrrrcrrr}
  \tabletypesize{\scriptsize}
  \tablenum{3}
  \tablecolumns{8}
  \tablewidth{0pc}
  \tablecaption{Emission Line Intensities and Equivalent Widths \label{tab3}}
\tablehead{
  \colhead{Ion}
  &{$F$($\lambda$)/$F$(H$\beta$)}
  &{$I$($\lambda$)/$I$(H$\beta$)}
  &{EW\tablenotemark{a}}&
  &{$F$($\lambda$)/$F$(H$\beta$)}
  &{$I$($\lambda$)/$I$(H$\beta$)}
  &{EW\tablenotemark{a}}
}
  \startdata
 &\multicolumn{7}{c}{\sc Galaxy}\\     \cline{2-8}  
 & \multicolumn{3}{c}{J0113+0052 No.1      }&&
 \multicolumn{3}{c}{J0113+0052 No.2      } \\ 
  \cline{2-4} \cline{6-8}
3727 [O {\sc ii}]                 &  50.41 $\pm$   1.88 &  78.78 $\pm$   3.02 &  54.1 & & 203.01 $\pm$   6.89 & 187.30 $\pm$   7.26 &  72.1 \\
3835 H9                           &   6.83 $\pm$   1.10 &  10.17 $\pm$   1.96 &   8.1 & &  \nodata~~~~~       &  \nodata~~~~~       &\nodata\\
3868 [Ne {\sc iii}]               &  12.60 $\pm$   1.06 &  18.49 $\pm$   1.58 &  11.2 & &  \nodata~~~~~       &  \nodata~~~~~       &\nodata\\
3889 He {\sc i} + H8              &  13.18 $\pm$   0.91 &  19.19 $\pm$   1.83 &  12.9 & &  \nodata~~~~~       &  \nodata~~~~~       &\nodata\\
3968 [Ne {\sc iii}] + H7          &  13.89 $\pm$   0.87 &  19.54 $\pm$   1.67 &  14.7 & &  \nodata~~~~~       &  \nodata~~~~~       &\nodata\\
4101 H$\delta$                    &  20.73 $\pm$   1.03 &  27.58 $\pm$   1.83 &  19.4 & &  19.57 $\pm$   1.74 &  27.81 $\pm$   3.30 &   7.3 \\
4340 H$\gamma$                    &  38.70 $\pm$   1.21 &  46.82 $\pm$   1.88 &  33.8 & &  41.30 $\pm$   2.16 &  46.74 $\pm$   3.04 &  17.4 \\
4363 [O {\sc iii}]                &   5.75 $\pm$   0.60 &   6.90 $\pm$   0.73 &   4.3 & &  \nodata~~~~~       &  \nodata~~~~~       &\nodata\\
4471 He {\sc i}                   &   3.36 $\pm$   0.55 &   3.86 $\pm$   0.64 &   2.7 & &  \nodata~~~~~       &  \nodata~~~~~       &\nodata\\
4861 H$\beta$                     & 100.00 $\pm$   2.34 & 100.00 $\pm$   2.57 &  83.4 & & 100.00 $\pm$   3.38 & 100.00 $\pm$   3.88 &  47.1 \\
4959 [O {\sc iii}]                &  73.86 $\pm$   1.85 &  71.46 $\pm$   1.81 &  60.5 & &  46.75 $\pm$   2.06 &  43.14 $\pm$   2.06 &  17.1 \\
5007 [O {\sc iii}]                & 212.88 $\pm$   4.54 & 202.67 $\pm$   4.37 & 194.6 & & 132.55 $\pm$   4.25 & 122.30 $\pm$   4.25 &  49.4 \\
 $C$(H$\beta$) & \multicolumn{3}{c}{ 0.630 }& & \multicolumn{3}{c}{ 0.000 } \\
 $F$(H$\beta$)\tablenotemark{b} & \multicolumn{3}{c}{  0.07 }& & \multicolumn{3}{c}{  0.03 } \\
 EW(abs) \AA & \multicolumn{3}{c}{ 0.00 }& & \multicolumn{3}{c}{ 3.95 } \\
\tableline
 &\multicolumn{7}{c}{\sc Galaxy}\\     \cline{2-8}  
 & \multicolumn{3}{c}{J0113+0052 No.4      }&&
 \multicolumn{3}{c}{J0204-1009 No.1      } \\ 
  \cline{2-4} \cline{6-8}
3727 [O {\sc ii}]                 & 112.24 $\pm$   4.56 & 107.64 $\pm$   4.76 &  75.9 & & 213.20 $\pm$  11.39 & 164.95 $\pm$  12.05 &   9.2 \\
3868 [Ne {\sc iii}]               &  14.09 $\pm$   1.92 &  13.51 $\pm$   1.93 &  13.8 & &  27.71 $\pm$   5.19 &  21.44 $\pm$   5.21 &   0.9 \\
4340 H$\gamma$                    &  41.84 $\pm$   2.00 &  47.36 $\pm$   4.05 &  33.0 & &  36.23 $\pm$   3.63 &  55.14 $\pm$   9.01 &   2.1 \\
4861 H$\beta$                     & 100.00 $\pm$   3.59 & 100.00 $\pm$   4.18 & 139.3 & & 100.00 $\pm$   5.49 & 100.00 $\pm$   7.87 &   6.8 \\
4959 [O {\sc iii}]                &  73.08 $\pm$   2.94 &  70.09 $\pm$   2.94 &  83.5 & &  64.56 $\pm$   4.45 &  49.95 $\pm$   4.45 &   3.4 \\
5007 [O {\sc iii}]                & 210.48 $\pm$   6.37 & 201.86 $\pm$   6.37 & 149.2 & & 205.73 $\pm$   9.53 & 159.17 $\pm$   9.53 &  11.4 \\
 $C$(H$\beta$) & \multicolumn{3}{c}{ 0.000 }& & \multicolumn{3}{c}{ 0.000 } \\
 $F$(H$\beta$)\tablenotemark{b} & \multicolumn{3}{c}{  0.03 }& & \multicolumn{3}{c}{  0.03 } \\
 EW(abs) \AA & \multicolumn{3}{c}{ 5.95 }& & \multicolumn{3}{c}{ 2.00 } \\ \tableline
 &\multicolumn{7}{c}{\sc Galaxy}\\     \cline{2-8}  
 & \multicolumn{3}{c}{J0204-1009 No.2      }&&
 \multicolumn{3}{c}{J0254+0035         } \\ 
  \cline{2-4} \cline{6-8}
3727 [O {\sc ii}]                 & 139.72 $\pm$   2.95 & 139.72 $\pm$   3.09 &  63.3 & & 166.40 $\pm$   6.42 & 153.25 $\pm$   6.91 &  20.6 \\
3750 H12                          &   2.01 $\pm$   0.53 &   2.01 $\pm$   0.81 &   1.0 & &  \nodata~~~~~       &  \nodata~~~~~       &\nodata\\
3771 H11                          &   3.15 $\pm$   0.70 &   3.15 $\pm$   0.93 &   1.6 & &  \nodata~~~~~       &  \nodata~~~~~       &\nodata\\
3798 H10                          &   4.72 $\pm$   0.61 &   4.72 $\pm$   0.86 &   2.4 & &  \nodata~~~~~       &  \nodata~~~~~       &\nodata\\
3835 H9                           &   6.32 $\pm$   0.78 &   6.32 $\pm$   1.00 &   3.1 & &  \nodata~~~~~       &  \nodata~~~~~       &\nodata\\
3868 [Ne {\sc iii}]               &  23.40 $\pm$   0.90 &  23.40 $\pm$   0.91 &   9.3 & &  10.44 $\pm$   2.09 &   9.59 $\pm$   2.14 &   1.0 \\
3889 He {\sc i} + H8              &  19.66 $\pm$   0.85 &  19.66 $\pm$   1.06 &   9.9 & &  11.48 $\pm$   1.74 &  23.95 $\pm$   5.47 &   1.6 \\
3968 [Ne {\sc iii}] + H7          &  22.84 $\pm$   0.99 &  22.84 $\pm$   1.17 &  11.7 & &   5.32 $\pm$   1.24 &  19.04 $\pm$   8.45 &   0.7 \\
4101 H$\delta$                    &  28.84 $\pm$   0.95 &  28.84 $\pm$   1.08 &  18.1 & &  15.05 $\pm$   2.19 &  27.36 $\pm$   5.35 &   2.0 \\
4340 H$\gamma$                    &  46.82 $\pm$   1.11 &  46.82 $\pm$   1.23 &  28.0 & &  38.05 $\pm$   2.36 &  46.74 $\pm$   3.65 &   5.7 \\
4363 [O {\sc iii}]                &   6.53 $\pm$   0.51 &   6.53 $\pm$   0.52 &   3.5 & &  \nodata~~~~~       &  \nodata~~~~~       &\nodata\\
4471 He {\sc i}                   &   3.70 $\pm$   0.58 &   3.70 $\pm$   0.58 &   2.0 & &  \nodata~~~~~       &  \nodata~~~~~       &\nodata\\
4861 H$\beta$                     & 100.00 $\pm$   1.94 & 100.00 $\pm$   1.97 &  83.4 & & 100.00 $\pm$   3.71 & 100.00 $\pm$   4.30 &  17.7 \\
4959 [O {\sc iii}]                &  94.11 $\pm$   1.86 &  94.11 $\pm$   1.86 &  76.5 & &  45.99 $\pm$   2.57 &  41.24 $\pm$   2.57 &   6.5 \\
5007 [O {\sc iii}]                & 272.63 $\pm$   4.86 & 272.63 $\pm$   4.86 & 222.6 & & 143.69 $\pm$   5.08 & 128.73 $\pm$   5.07 &  21.5 \\
5015 He {\sc i}                   &   1.97 $\pm$   0.31 &   1.97 $\pm$   0.31 &   1.6 & &  \nodata~~~~~       &  \nodata~~~~~       &\nodata\\
 $C$(H$\beta$) & \multicolumn{3}{c}{ 0.000 }& & \multicolumn{3}{c}{ 0.035 } \\
 $F$(H$\beta$)\tablenotemark{b} & \multicolumn{3}{c}{  0.20 }& & \multicolumn{3}{c}{  0.03 } \\
 EW(abs) \AA & \multicolumn{3}{c}{ 0.00 }& & \multicolumn{3}{c}{ 2.00 } \\ \tableline
 &\multicolumn{7}{c}{\sc Galaxy}\\     \cline{2-8}  
 & \multicolumn{3}{c}{J0301-0052         }&&
 \multicolumn{3}{c}{J0313+0010         } \\ 
  \cline{2-4} \cline{6-8}
3727 [O {\sc ii}]                 &  45.17 $\pm$   2.48 &  59.92 $\pm$   3.55 &  37.3 & & 243.68 $\pm$  12.33 & 203.20 $\pm$  13.13 &  19.8 \\
3868 [Ne {\sc iii}]               &  23.35 $\pm$   1.73 &  29.55 $\pm$   2.34 &  17.8 & &  \nodata~~~~~       &  \nodata~~~~~       &\nodata\\
3889 He {\sc i} + H8              &  12.20 $\pm$   1.62 &  21.42 $\pm$   3.36 &   9.2 & &  \nodata~~~~~       &  \nodata~~~~~       &\nodata\\
3968 [Ne {\sc iii}] + H7          &  15.39 $\pm$   1.78 &  25.59 $\pm$   3.52 &  10.2 & &  \nodata~~~~~       &  \nodata~~~~~       &\nodata\\
4101 H$\delta$                    &  20.49 $\pm$   1.74 &  27.12 $\pm$   2.54 &  28.9 & &  \nodata~~~~~       &  \nodata~~~~~       &\nodata\\
4340 H$\gamma$                    &  37.00 $\pm$   1.11 &  47.13 $\pm$   2.07 &  22.1 & &  35.04 $\pm$   3.52 &  49.12 $\pm$   7.56 &   2.9 \\
4363 [O {\sc iii}]                &   8.63 $\pm$   1.00 &   9.38 $\pm$   1.15 &   5.1 & &  \nodata~~~~~       &  \nodata~~~~~       &\nodata\\
4861 H$\beta$                     & 100.00 $\pm$   2.84 & 100.00 $\pm$   3.22 &  65.7 & & 100.00 $\pm$   5.54 & 100.00 $\pm$   7.51 &  10.0 \\
4959 [O {\sc iii}]                & 132.88 $\pm$   3.53 & 122.78 $\pm$   3.47 &  82.1 & &  64.61 $\pm$   4.46 &  53.88 $\pm$   4.46 &   5.1 \\
5007 [O {\sc iii}]                & 395.54 $\pm$   9.32 & 361.08 $\pm$   9.06 & 230.6 & & 186.08 $\pm$   8.79 & 155.17 $\pm$   8.80 &  14.2 \\
 $C$(H$\beta$) & \multicolumn{3}{c}{ 0.475 }& & \multicolumn{3}{c}{ 0.000 } \\
 $F$(H$\beta$)\tablenotemark{b} & \multicolumn{3}{c}{  0.06 }& & \multicolumn{3}{c}{  0.02 } \\
 EW(abs) \AA & \multicolumn{3}{c}{ 3.65 }& & \multicolumn{3}{c}{ 2.00 } \\ \tableline
 &\multicolumn{7}{c}{\sc Galaxy}\\     \cline{2-8}  
 & \multicolumn{3}{c}{J0747+5111 No.1      }&&
 \multicolumn{3}{c}{J0747+5111 No.2      } \\ 
  \cline{2-4} \cline{6-8}
3727 [O {\sc ii}]                 & 195.99 $\pm$   3.92 & 240.79 $\pm$   5.27 &  37.2 & & 157.69 $\pm$   2.89 & 160.82 $\pm$   3.13 &  75.9 \\
3771 H11                          &  \nodata~~~~~       &  \nodata~~~~~       &\nodata& &   2.44 $\pm$   0.54 &   4.99 $\pm$   1.37 &   1.2 \\
3798 H10                          &  \nodata~~~~~       &  \nodata~~~~~       &\nodata& &   4.24 $\pm$   0.56 &   6.74 $\pm$   1.10 &   2.2 \\
3835 H9                           &  \nodata~~~~~       &  \nodata~~~~~       &\nodata& &   5.24 $\pm$   0.48 &   7.70 $\pm$   0.93 &   2.8 \\
3868 [Ne {\sc iii}]               &  29.31 $\pm$   1.38 &  34.79 $\pm$   1.73 &   4.6 & &  32.59 $\pm$   0.93 &  33.09 $\pm$   0.97 &  16.8 \\
3889 He {\sc i} + H8              &  14.77 $\pm$   1.07 &  23.37 $\pm$   2.12 &   3.1 & &  17.65 $\pm$   0.67 &  20.24 $\pm$   0.95 &   9.6 \\
3968 [Ne {\sc iii}] + H7          &  15.41 $\pm$   1.17 &  24.06 $\pm$   2.27 &   3.0 & &  24.20 $\pm$   0.76 &  26.80 $\pm$   1.01 &  13.3 \\
4101 H$\delta$                    &  21.60 $\pm$   1.07 &  29.86 $\pm$   1.87 &   4.6 & &  23.87 $\pm$   0.67 &  26.06 $\pm$   0.87 &  15.0 \\
4340 H$\gamma$                    &  39.29 $\pm$   1.11 &  47.11 $\pm$   1.69 &   8.6 & &  45.67 $\pm$   0.93 &  47.31 $\pm$   1.05 &  36.3 \\
4363 [O {\sc iii}]                &   7.29 $\pm$   0.76 &   7.74 $\pm$   0.84 &   1.3 & &   8.14 $\pm$   0.43 &   8.14 $\pm$   0.44 &   5.3 \\
4471 He {\sc i}                   &   2.07 $\pm$   0.50 &   2.15 $\pm$   0.55 &   0.4 & &   3.37 $\pm$   0.36 &   3.36 $\pm$   0.37 &   2.4 \\
4686 He {\sc ii}                  &  \nodata~~~~~       &  \nodata~~~~~       &\nodata& &   0.89 $\pm$   0.22 &   0.89 $\pm$   0.23 &   0.7 \\
4861 H$\beta$                     & 100.00 $\pm$   1.93 & 100.00 $\pm$   2.14 &  26.6 & & 100.00 $\pm$   1.73 & 100.00 $\pm$   1.77 & 101.4 \\
4959 [O {\sc iii}]                & 125.97 $\pm$   2.35 & 119.01 $\pm$   2.32 &  27.4 & & 129.79 $\pm$   2.19 & 127.91 $\pm$   2.19 & 122.6 \\
5007 [O {\sc iii}]                & 381.73 $\pm$   6.63 & 357.49 $\pm$   6.49 &  83.1 & & 387.31 $\pm$   6.24 & 381.25 $\pm$   6.23 & 358.1 \\
5015 He {\sc i}                   &  \nodata~~~~~       &  \nodata~~~~~       &\nodata& &   2.42 $\pm$   0.23 &   2.38 $\pm$   0.23 &   2.3 \\
 $C$(H$\beta$) & \multicolumn{3}{c}{ 0.345 }& & \multicolumn{3}{c}{ 0.045 } \\
 $F$(H$\beta$)\tablenotemark{b} & \multicolumn{3}{c}{  0.28 }& & \multicolumn{3}{c}{  0.29 } \\
 EW(abs) \AA & \multicolumn{3}{c}{ 1.05 }& & \multicolumn{3}{c}{ 1.25 } \\ \tableline
 &\multicolumn{7}{c}{\sc Galaxy}\\     \cline{2-8}  
 &  \multicolumn{3}{c}{J0812+4836         }&&
 \multicolumn{3}{c}{J0859+3923         } \\ 
  \cline{2-4} \cline{6-8}
3727 [O {\sc ii}]                 & 161.92 $\pm$   5.47 & 194.96 $\pm$   7.50 &  31.6 & & 307.16 $\pm$  11.73 & 276.72 $\pm$  12.51 &  41.5 \\
3868 [Ne {\sc iii}]               &   5.64 $\pm$   1.38 &   6.54 $\pm$   1.75 &   0.8 & &  18.41 $\pm$   3.37 &  16.59 $\pm$   3.37 &   1.7 \\
3889 He {\sc i} + H8              &   6.22 $\pm$   2.35 &  20.45 $\pm$   9.64 &   1.2 & &  \nodata~~~~~       &  \nodata~~~~~       &\nodata\\
3968 [Ne {\sc iii}] + H7          &   8.79 $\pm$   1.85 &  20.59 $\pm$   5.50 &   2.0 & &  \nodata~~~~~       &  \nodata~~~~~       &\nodata\\
4101 H$\delta$                    &  14.15 $\pm$   1.91 &  27.21 $\pm$   4.82 &   2.8 & &  17.10 $\pm$   2.42 &  27.17 $\pm$   5.25 &   2.2 \\
4340 H$\gamma$                    &  34.67 $\pm$   2.04 &  46.60 $\pm$   3.59 &   7.2 & &  39.76 $\pm$   2.72 &  46.74 $\pm$   4.14 &   5.4 \\
4861 H$\beta$                     & 100.00 $\pm$   3.25 & 100.00 $\pm$   3.79 &  25.8 & & 100.00 $\pm$   4.23 & 100.00 $\pm$   5.03 &  15.0 \\
4959 [O {\sc iii}]                &  28.99 $\pm$   1.71 &  26.23 $\pm$   1.69 &   6.2 & &  56.49 $\pm$   2.98 &  50.89 $\pm$   2.98 &   6.9 \\
5007 [O {\sc iii}]                &  93.25 $\pm$   3.09 &  83.59 $\pm$   3.02 &  19.8 & & 181.47 $\pm$   6.75 & 163.49 $\pm$   6.76 &  22.5 \\
 $C$(H$\beta$) & \multicolumn{3}{c}{ 0.375 }& & \multicolumn{3}{c}{ 0.000 } \\
 $F$(H$\beta$)\tablenotemark{b} & \multicolumn{3}{c}{  0.07 }& & \multicolumn{3}{c}{  0.03 } \\
 EW(abs) \AA & \multicolumn{3}{c}{ 2.15 }& & \multicolumn{3}{c}{ 1.65 } \\ \tableline
 &\multicolumn{7}{c}{\sc Galaxy}\\     \cline{2-8}  
 &  \multicolumn{3}{c}{J0911+3135         }&&
 \multicolumn{3}{c}{J0940+2935         } \\ 
  \cline{2-4} \cline{6-8}
3727 [O {\sc ii}]                 & 219.22 $\pm$   7.75 & 279.68 $\pm$  10.95 &  21.9 & & 165.62 $\pm$   3.23 & 185.55 $\pm$   3.93 &  39.9 \\
3835 H9                           &  \nodata~~~~~       &  \nodata~~~~~       &\nodata& &   2.99 $\pm$   0.68 &   8.49 $\pm$   2.46 &   0.8 \\
3868 [Ne {\sc iii}]               &  10.11 $\pm$   3.28 &  12.39 $\pm$   4.23 &   0.8 & &  14.33 $\pm$   0.90 &  15.73 $\pm$   1.03 &   3.2 \\
3889 He {\sc i} + H8              &  10.00 $\pm$   1.81 &  19.05 $\pm$   5.07 &   1.1 & &  12.69 $\pm$   0.76 &  19.20 $\pm$   1.55 &   3.4 \\
3968 [Ne {\sc iii}] + H7          &   6.38 $\pm$   1.16 &  13.99 $\pm$   4.52 &   0.7 & &  13.64 $\pm$   0.77 &  19.98 $\pm$   1.52 &   3.7 \\
4101 H$\delta$                    &  22.77 $\pm$   1.68 &  32.22 $\pm$   3.63 &   2.6 & &  20.10 $\pm$   0.85 &  26.18 $\pm$   1.44 &   5.8 \\
4340 H$\gamma$                    &  38.21 $\pm$   2.28 &  47.13 $\pm$   3.86 &   4.3 & &  41.81 $\pm$   1.02 &  47.12 $\pm$   1.40 &  13.9 \\
4363 [O {\sc iii}]                &   2.13 $\pm$   0.67 &   2.29 $\pm$   0.76 &   0.2 & &   4.13 $\pm$   0.54 &   4.25 $\pm$   0.58 &   1.2 \\
4471 He {\sc i}                   &  \nodata~~~~~       &  \nodata~~~~~       &\nodata& &   2.14 $\pm$   0.51 &   2.17 $\pm$   0.53 &   0.6 \\
4861 H$\beta$                     & 100.00 $\pm$   3.68 & 100.00 $\pm$   4.27 &  13.5 & & 100.00 $\pm$   1.87 & 100.00 $\pm$   2.02 &  40.4 \\
4959 [O {\sc iii}]                &  35.14 $\pm$   2.36 &  32.93 $\pm$   2.32 &   4.2 & &  77.75 $\pm$   1.52 &  74.51 $\pm$   1.51 &  29.5 \\
5007 [O {\sc iii}]                & 101.15 $\pm$   3.77 &  93.83 $\pm$   3.67 &  12.0 & & 234.50 $\pm$   3.91 & 223.58 $\pm$   3.86 &  89.0 \\
5015 He {\sc i}                   &  \nodata~~~~~       &  \nodata~~~~~       &\nodata& &   1.75 $\pm$   0.03 &   1.66 $\pm$   0.03 &   0.7 \\
 $C$(H$\beta$) & \multicolumn{3}{c}{ 0.405 }& & \multicolumn{3}{c}{ 0.205 } \\
 $F$(H$\beta$)\tablenotemark{b} & \multicolumn{3}{c}{  0.05 }& & \multicolumn{3}{c}{  0.34 } \\
 EW(abs) \AA & \multicolumn{3}{c}{ 0.60 }& & \multicolumn{3}{c}{ 1.30 } \\ \tableline
 &\multicolumn{7}{c}{\sc Galaxy}\\     \cline{2-8}  
 &  \multicolumn{3}{c}{J0946+5452         }&&
 \multicolumn{3}{c}{J0956+2849 No.1      } \\ 
  \cline{2-4} \cline{6-8}
3727 [O {\sc ii}]                 & 178.39 $\pm$   5.59 & 208.41 $\pm$   6.99 &  29.9 & &  30.46 $\pm$   0.81 &  39.18 $\pm$   1.08 &  41.5 \\
3750 H12                          &  \nodata~~~~~       &  \nodata~~~~~       &\nodata& &   2.29 $\pm$   0.34 &   4.77 $\pm$   0.90 &   3.2 \\
3771 H11                          &  \nodata~~~~~       &  \nodata~~~~~       &\nodata& &   2.37 $\pm$   0.35 &   4.68 $\pm$   0.85 &   3.6 \\
3798 H10                          &  \nodata~~~~~       &  \nodata~~~~~       &\nodata& &   3.18 $\pm$   0.36 &   5.73 $\pm$   0.80 &   4.7 \\
3835 H9                           &  \nodata~~~~~       &  \nodata~~~~~       &\nodata& &   5.04 $\pm$   0.36 &   8.09 $\pm$   0.75 &   7.0 \\
3868 [Ne {\sc iii}]               &  27.44 $\pm$   2.24 &  31.29 $\pm$   2.62 &   3.3 & &  10.28 $\pm$   0.44 &  12.74 $\pm$   0.56 &  13.6 \\
3889 He {\sc i} + H8              &  10.65 $\pm$   1.88 &  14.74 $\pm$   3.21 &   1.8 & &  14.35 $\pm$   0.45 &  19.44 $\pm$   0.76 &  20.3 \\
3968 [Ne {\sc iii}] + H7          &  11.55 $\pm$   1.44 &  15.63 $\pm$   2.70 &   1.9 & &  15.54 $\pm$   0.66 &  20.56 $\pm$   1.00 &  21.3 \\
4101 H$\delta$                    &  19.65 $\pm$   1.75 &  24.11 $\pm$   2.78 &   3.4 & &  20.29 $\pm$   0.50 &  25.39 $\pm$   0.75 &  28.9 \\
4340 H$\gamma$                    &  42.38 $\pm$   1.98 &  47.19 $\pm$   2.75 &   7.5 & &  41.58 $\pm$   0.79 &  47.53 $\pm$   1.00 &  60.5 \\
4363 [O {\sc iii}]                &   7.72 $\pm$   1.63 &   8.14 $\pm$   1.75 &   1.0 & &   5.42 $\pm$   0.30 &   5.97 $\pm$   0.34 &   8.2 \\
4471 He {\sc i}                   &  \nodata~~~~~       &  \nodata~~~~~       &\nodata& &   3.03 $\pm$   0.28 &   3.25 $\pm$   0.31 &   4.8 \\
4686 He {\sc ii}                  &  \nodata~~~~~       &  \nodata~~~~~       &\nodata& &   2.10 $\pm$   0.24 &   2.15 $\pm$   0.25 &   3.3 \\
4861 H$\beta$                     & 100.00 $\pm$   3.03 & 100.00 $\pm$   3.32 &  21.9 & & 100.00 $\pm$   1.65 & 100.00 $\pm$   1.70 & 187.9 \\
4959 [O {\sc iii}]                & 122.28 $\pm$   3.56 & 118.55 $\pm$   3.53 &  20.2 & &  62.78 $\pm$   1.08 &  60.92 $\pm$   1.06 & 109.7 \\
5007 [O {\sc iii}]                & 371.77 $\pm$   9.26 & 358.18 $\pm$   9.13 &  63.1 & & 189.44 $\pm$   3.02 & 182.12 $\pm$   2.95 & 326.5 \\
 $C$(H$\beta$) & \multicolumn{3}{c}{ 0.245 }& & \multicolumn{3}{c}{ 0.370 } \\
 $F$(H$\beta$)\tablenotemark{b} & \multicolumn{3}{c}{  0.09 }& & \multicolumn{3}{c}{  0.34 } \\
 EW(abs) \AA & \multicolumn{3}{c}{ 0.40 }& & \multicolumn{3}{c}{ 2.00 } \\ \tableline
 &\multicolumn{7}{c}{\sc Galaxy}\\     \cline{2-8}  
 &  \multicolumn{3}{c}{J0956+2849 No.2      }&&
 \multicolumn{3}{c}{J0956+2849 No.6      } \\ 
  \cline{2-4} \cline{6-8}
3727 [O {\sc ii}]                 &  54.36 $\pm$   2.11 &  68.83 $\pm$   2.76 &  71.9 & & 125.24 $\pm$   5.57 & 138.77 $\pm$   6.69 &  44.6 \\
3771 H11                          &   3.49 $\pm$   0.99 &   6.51 $\pm$   2.17 &   7.2 & &  \nodata~~~~~       &  \nodata~~~~~       &\nodata\\
3798 H10                          &   4.97 $\pm$   0.89 &   8.74 $\pm$   2.03 &   8.6 & &  \nodata~~~~~       &  \nodata~~~~~       &\nodata\\
3835 H9                           &   6.75 $\pm$   1.21 &  10.67 $\pm$   2.23 &  12.6 & &  \nodata~~~~~       &  \nodata~~~~~       &\nodata\\
3868 [Ne {\sc iii}]               &   8.45 $\pm$   1.15 &  10.34 $\pm$   1.43 &  14.3 & &   8.62 $\pm$   2.80 &   9.36 $\pm$   3.19 &   2.6 \\
3889 He {\sc i} + H8              &  14.92 $\pm$   1.90 &  20.15 $\pm$   2.77 &  32.4 & &   9.36 $\pm$   1.90 &  15.19 $\pm$   4.22 &   4.0 \\
3968 [Ne {\sc iii}] + H7          &  15.61 $\pm$   1.01 &  21.38 $\pm$   1.89 &  24.2 & &   6.88 $\pm$   1.90 &  13.68 $\pm$   5.44 &   2.3 \\
4101 H$\delta$                    &  20.06 $\pm$   1.03 &  25.75 $\pm$   1.77 &  32.7 & &  21.54 $\pm$   2.53 &  27.84 $\pm$   4.14 &   8.7 \\
4340 H$\gamma$                    &  41.50 $\pm$   1.25 &  47.48 $\pm$   1.69 &  85.1 & &  41.51 $\pm$   2.43 &  47.62 $\pm$   3.74 &  15.9 \\
4363 [O {\sc iii}]                &   4.04 $\pm$   0.72 &   4.41 $\pm$   0.80 &   9.0 & &   5.56 $\pm$   1.65 &   5.66 $\pm$   1.76 &   1.7 \\
4471 He {\sc i}                   &   3.30 $\pm$   0.59 &   3.52 $\pm$   0.65 &   7.6 & &   5.37 $\pm$   1.89 &   5.39 $\pm$   1.99 &   1.9 \\
4686 He {\sc ii}                  &   6.19 $\pm$   0.76 &   6.32 $\pm$   0.79 &  16.4 & &  \nodata~~~~~       &  \nodata~~~~~       &\nodata\\
4861 H$\beta$                     & 100.00 $\pm$   2.28 & 100.00 $\pm$   2.38 & 294.2 & & 100.00 $\pm$   3.60 & 100.00 $\pm$   4.18 &  45.8 \\
4959 [O {\sc iii}]                &  44.81 $\pm$   1.26 &  43.47 $\pm$   1.24 & 145.0 & &  48.67 $\pm$   2.25 &  46.14 $\pm$   2.23 &  20.0 \\
5007 [O {\sc iii}]                & 138.05 $\pm$   2.96 & 132.73 $\pm$   2.90 & 540.1 & & 153.05 $\pm$   5.01 & 144.31 $\pm$   4.95 &  61.7 \\
 $C$(H$\beta$) & \multicolumn{3}{c}{ 0.350 }& & \multicolumn{3}{c}{ 0.205 } \\
 $F$(H$\beta$)\tablenotemark{b} & \multicolumn{3}{c}{  0.11 }& & \multicolumn{3}{c}{  0.04 } \\
 EW(abs) \AA & \multicolumn{3}{c}{ 3.55 }& & \multicolumn{3}{c}{ 2.00 } \\ \tableline
 &\multicolumn{7}{c}{\sc Galaxy}\\     \cline{2-8}  
 &  \multicolumn{3}{c}{J2238+1400 No.1      }&&
 \multicolumn{3}{c}{J2238+1400 No.2      } \\ 
  \cline{2-4} \cline{6-8}
3188 He {\sc i}                   &   1.51 $\pm$   0.22 &   2.04 $\pm$   0.30 &   2.9 & &  \nodata~~~~~       &  \nodata~~~~~       &\nodata\\
3679 H21                          &   0.43 $\pm$   0.07 &   0.52 $\pm$   0.09 &   0.9 & &  \nodata~~~~~       &  \nodata~~~~~       &\nodata\\
3683 H20                          &   0.55 $\pm$   0.10 &   0.67 $\pm$   0.12 &   1.2 & &   0.86 $\pm$   0.18 &   1.23 $\pm$   0.26 &   1.3 \\
3687 H19                          &   0.68 $\pm$   0.10 &   0.83 $\pm$   0.13 &   1.5 & &   0.65 $\pm$   0.13 &   0.94 $\pm$   0.19 &   1.0 \\
3692 H18                          &   1.05 $\pm$   0.11 &   1.28 $\pm$   0.13 &   2.4 & &   0.72 $\pm$   0.13 &   1.03 $\pm$   0.18 &   1.1 \\
3697 H17                          &   1.05 $\pm$   0.11 &   1.27 $\pm$   0.13 &   2.4 & &   0.72 $\pm$   0.13 &   1.03 $\pm$   0.18 &   1.1 \\
3704 H16                          &   1.53 $\pm$   0.11 &   1.86 $\pm$   0.14 &   3.5 & &   1.35 $\pm$   0.17 &   1.93 $\pm$   0.24 &   2.0 \\
3712 H15                          &   1.40 $\pm$   0.10 &   2.03 $\pm$   0.20 &   3.2 & &   1.28 $\pm$   0.22 &   1.83 $\pm$   0.38 &   2.0 \\
3722 H14                          &   1.86 $\pm$   0.09 &   2.25 $\pm$   0.11 &   4.3 & &   1.18 $\pm$   0.09 &   1.67 $\pm$   0.13 &   1.9 \\
3727 [O {\sc ii}]                 &  25.39 $\pm$   0.41 &  30.68 $\pm$   0.52 &  59.0 & &  40.33 $\pm$   0.69 &  57.08 $\pm$   1.03 &  63.1 \\
3750 H12                          &   2.59 $\pm$   0.12 &   3.45 $\pm$   0.21 &   6.0 & &   2.35 $\pm$   0.18 &   3.29 $\pm$   0.34 &   3.6 \\
3771 H11                          &   3.18 $\pm$   0.12 &   4.15 $\pm$   0.21 &   7.3 & &   2.38 $\pm$   0.17 &   3.31 $\pm$   0.32 &   3.6 \\
3798 H10                          &   4.50 $\pm$   0.14 &   5.70 $\pm$   0.23 &  10.3 & &   3.81 $\pm$   0.18 &   5.26 $\pm$   0.33 &   5.9 \\
3820 He {\sc i}                   &   1.07 $\pm$   0.10 &   1.27 $\pm$   0.12 &   2.5 & &   1.07 $\pm$   0.14 &   1.46 $\pm$   0.20 &   1.6 \\
3835 H9                           &   5.96 $\pm$   0.15 &   7.40 $\pm$   0.23 &  13.5 & &   4.87 $\pm$   0.20 &   6.64 $\pm$   0.34 &   7.5 \\
3868 [Ne {\sc iii}]               &  35.74 $\pm$   0.55 &  42.05 $\pm$   0.68 &  80.5 & &  30.17 $\pm$   0.54 &  40.67 $\pm$   0.75 &  45.3 \\
3889 He {\sc i} + H8              &  14.41 $\pm$   0.25 &  17.23 $\pm$   0.34 &  32.5 & &  14.26 $\pm$   0.32 &  19.09 $\pm$   0.49 &  21.0 \\
3968 [Ne {\sc iii}] + H7          &  25.23 $\pm$   0.40 &  29.46 $\pm$   0.50 &  58.6 & &  20.78 $\pm$   0.40 &  27.10 $\pm$   0.57 &  30.9 \\
4026 He {\sc i}                   &   1.54 $\pm$   0.09 &   1.75 $\pm$   0.11 &   3.5 & &   1.57 $\pm$   0.17 &   2.01 $\pm$   0.22 &   2.4 \\
4101 H$\delta$                    &  23.15 $\pm$   0.36 &  26.40 $\pm$   0.44 &  58.1 & &  21.34 $\pm$   0.40 &  26.65 $\pm$   0.54 &  34.6 \\
4143 He {\sc i}                   &   0.35 $\pm$   0.05 &   0.39 $\pm$   0.06 &   0.8 & &  \nodata~~~~~       &  \nodata~~~~~       &\nodata\\
4227 [Fe {\sc v}]                 &   0.63 $\pm$   0.06 &   0.70 $\pm$   0.07 &   1.6 & &  \nodata~~~~~       &  \nodata~~~~~       &\nodata\\
4340 H$\gamma$                    &  43.61 $\pm$   0.65 &  47.48 $\pm$   0.73 & 112.8 & &  40.61 $\pm$   0.66 &  47.09 $\pm$   0.80 &  65.6 \\
4363 [O {\sc iii}]                &  16.72 $\pm$   0.27 &  18.04 $\pm$   0.30 &  43.2 & &  12.27 $\pm$   0.26 &  14.14 $\pm$   0.31 &  19.8 \\
4387 He {\sc i}                   &   0.42 $\pm$   0.06 &   0.46 $\pm$   0.07 &   1.1 & &  \nodata~~~~~       &  \nodata~~~~~       &\nodata\\
4471 He {\sc i}                   &   3.63 $\pm$   0.10 &   3.85 $\pm$   0.10 &   9.9 & &   3.61 $\pm$   0.17 &   4.02 $\pm$   0.19 &   6.2 \\
4658 [Fe {\sc iii}]               &   0.33 $\pm$   0.05 &   0.34 $\pm$   0.05 &   0.9 & &  \nodata~~~~~       &  \nodata~~~~~       &\nodata\\
4686 He {\sc ii}                  &   1.02 $\pm$   0.08 &   1.05 $\pm$   0.08 &   3.1 & &   1.06 $\pm$   0.13 &   1.12 $\pm$   0.13 &   2.0 \\
4711 [Ar {\sc iv}] + He {\sc i}   &   3.50 $\pm$   0.10 &   3.57 $\pm$   0.10 &  10.5 & &   2.03 $\pm$   0.13 &   2.12 $\pm$   0.14 &   3.7 \\
4740 [Ar {\sc iv}]                &   2.33 $\pm$   0.08 &   2.37 $\pm$   0.08 &   7.0 & &   1.37 $\pm$   0.13 &   1.42 $\pm$   0.14 &   2.7 \\
4861 H$\beta$                     & 100.00 $\pm$   1.46 & 100.00 $\pm$   1.47 & 303.9 & & 100.00 $\pm$   1.52 & 100.00 $\pm$   1.54 & 189.4 \\
4921 He {\sc i}                   &   0.91 $\pm$   0.07 &   0.90 $\pm$   0.07 &   2.8 & &   0.91 $\pm$   0.10 &   0.89 $\pm$   0.10 &   1.8 \\
4959 [O {\sc iii}]                & 168.98 $\pm$   2.45 & 166.24 $\pm$   2.43 & 510.5 & & 164.67 $\pm$   2.47 & 160.48 $\pm$   2.43 & 320.0 \\
4988 [Fe {\sc iii}]               &   0.58 $\pm$   0.07 &   0.57 $\pm$   0.07 &   1.7 & &  \nodata~~~~~       &  \nodata~~~~~       &\nodata\\
5007 [O {\sc iii}]                & 506.95 $\pm$   7.31 & 495.33 $\pm$   7.19 &1531.5 & & 494.02 $\pm$   7.33 & 475.50 $\pm$   7.11 & 960.0 \\
 $C$(H$\beta$) & \multicolumn{3}{c}{ 0.270 }& & \multicolumn{3}{c}{ 0.490 } \\
 $F$(H$\beta$)\tablenotemark{b} & \multicolumn{3}{c}{  2.52 }& & \multicolumn{3}{c}{  0.99 } \\
 EW(abs) \AA & \multicolumn{3}{c}{ 0.65 }& & \multicolumn{3}{c}{ 0.00 } \\
  \enddata
\tablenotetext{a}{in \AA.}
\tablenotetext{b}{in units 10$^{-14}$ erg s$^{-1}$ cm$^{-2}$.}
  \end{deluxetable}

\clearpage

  \begin{deluxetable}{lrrrr}
  \tablenum{4}
  \tabletypesize{\scriptsize}
  \tablecolumns{5}
  \tablewidth{0pc}
  \tablecaption{\sc Ionic and Total Heavy Element Abundances \label{tab4}}
  \tablehead{
  \colhead{\sc Property}&
\multicolumn{1}{c}{Value} & \multicolumn{1}{c}{Value} &
\multicolumn{1}{c}{Value} & \multicolumn{1}{c}{Value} }
  \startdata
 &\multicolumn{4}{c}{\sc Galaxy}\\      \cline{2-5}
 & \multicolumn{1}{c}{J0113+0052 No.1     } & \multicolumn{1}{c}{J0113+0052 No.2     } &
\multicolumn{1}{c}{J0113+0052 No.4     } & \multicolumn{1}{c}{J0204-1009 No.1     } \\ \tableline
  $T_{\rm e}$(O {\sc iii}) (K) &
$19963  \pm1290 $ & $18766  \pm1011 $ &
$18330  \pm1009 $ & $18417  \pm1032 $  \\

  $T_{\rm e}$(O {\sc ii}) (K) &
$16278  \pm\,~932 $ & $16014  \pm1255 $ &
$15883  \pm1244 $ & $15911  \pm1274 $  \\

  \\
  O$^+$/H$^+$ ($\times$10$^4$) &
$ 0.056 \pm 0.008$ & $ 0.139 \pm 0.027$ &
$ 0.082 \pm 0.016$ & $ 0.125 \pm 0.026$ \\
  O$^{++}$/H$^+$ ($\times$10$^4$) &
$ 0.118 \pm 0.017$ & $ 0.081 \pm 0.010$ &
$ 0.140 \pm 0.018$ & $ 0.107 \pm 0.015$ \\
  O/H ($\times$10$^4$) &
$ 0.174 \pm 0.019$ & $ 0.220 \pm 0.029$ &
$ 0.222 \pm 0.025$ & $ 0.232 \pm 0.030$ \\
  12 + log(O/H)  &
$ 7.240 \pm 0.048$ & $ 7.343 \pm 0.057$ &
$ 7.347 \pm 0.048$ & $ 7.365 \pm 0.057$ \\
  \\
  Ne$^{++}$/H$^+$ ($\times$10$^5$) &
$ 0.235 \pm 0.038$ &    \nodata~~~~~    &
$ 0.211 \pm 0.030$ & $ 0.331 \pm 0.081$ \\
  ICF &
  1.116 &  \nodata~~~~~ &
  1.139 &  1.239 \\
  log(Ne/O) &
$-0.821 \pm 0.104$ &    \nodata~~~~~    &
$-0.966 \pm 0.088$ & $-0.753 \pm 0.188$ \\ \tableline
  &\multicolumn{4}{c}{\sc Galaxy}\\      \cline{2-5}
 & \multicolumn{1}{c}{J0204-1009 No.2     } & \multicolumn{1}{c}{J0254+0035        } &
\multicolumn{1}{c}{J0301-0052        } & \multicolumn{1}{c}{J0313+0010        } 
 \\ \tableline
  $T_{\rm e}$(O {\sc iii}) (K) &
$16534  \pm\,~665 $ & $19285  \pm1013 $ &
$17257  \pm1125 $ & $17839  \pm1028 $  \\

  $T_{\rm e}$(O {\sc ii}) (K) &
$15148  \pm\,~553 $ & $16146  \pm1269 $ &
$15482  \pm\,~913 $ & $15714  \pm1257 $  \\

  \\
  O$^+$/H$^+$ ($\times$10$^4$) &
$ 0.123 \pm 0.012$ & $ 0.111 \pm 0.022$ &
$ 0.049 \pm 0.008$ & $ 0.160 \pm 0.034$ \\
  O$^{++}$/H$^+$ ($\times$10$^4$) &
$ 0.239 \pm 0.024$ & $ 0.079 \pm 0.010$ &
$ 0.285 \pm 0.045$ & $ 0.114 \pm 0.017$ \\
  O/H ($\times$10$^4$) &
$ 0.361 \pm 0.027$ & $ 0.190 \pm 0.024$ &
$ 0.335 \pm 0.046$ & $ 0.274 \pm 0.037$ \\
  12 + log(O/H)  &
$ 7.558 \pm 0.032$ & $ 7.279 \pm 0.055$ &
$ 7.524 \pm 0.060$ & $ 7.438 \pm 0.059$ \\
  \\
  Ne$^{++}$/H$^+$ ($\times$10$^5$) &
$ 0.476 \pm 0.050$ & $ 0.132 \pm 0.033$ &
$ 0.537 \pm 0.093$ &    \nodata~~~~~    \\
  ICF &
  1.125 &  1.271 &
  1.048 &  \nodata~~~~~ \\
  log(Ne/O) &
$-0.829 \pm 0.070$ & $-1.053 \pm 0.214$ &
$-0.774 \pm 0.103$ &    \nodata~~~~~    \\ \tableline
  &\multicolumn{4}{c}{\sc Galaxy}\\      \cline{2-5}
  & \multicolumn{1}{c}{J0747+5111 No.1     } & \multicolumn{1}{c}{J0747+5111 No.2     } &
\multicolumn{1}{c}{J0812+4836        } & \multicolumn{1}{c}{J0859+3923        }  \\ \tableline
  $T_{\rm e}$(O {\sc iii}) (K) &
$15816  \pm\,~832 $ & $15694  \pm\,~412 $ &
$19634  \pm1013 $ & $16799  \pm1015 $  \\

  $T_{\rm e}$(O {\sc ii}) (K) &
$14765  \pm\,~708 $ & $14695  \pm\,~352 $ &
$16220  \pm1276 $ & $15276  \pm1221 $  \\

  \\
  O$^+$/H$^+$ ($\times$10$^4$) &
$ 0.229 \pm 0.030$ & $ 0.155 \pm 0.010$ &
$ 0.140 \pm 0.027$ & $ 0.237 \pm 0.049$ \\
  O$^{++}$/H$^+$ ($\times$10$^4$) &
$ 0.345 \pm 0.046$ & $ 0.375 \pm 0.025$ &
$ 0.049 \pm 0.006$ & $ 0.134 \pm 0.020$ \\
  O$^{+++}$/H$^+$ ($\times$10$^6$) &
   \nodata~~~~~    & $ 0.562 \pm 0.146$ &
   \nodata~~~~~    &    \nodata~~~~~    \\
  O/H ($\times$10$^4$) &
$ 0.574 \pm 0.055$ & $ 0.536 \pm 0.027$ &
$ 0.189 \pm 0.028$ & $ 0.372 \pm 0.054$ \\
  12 + log(O/H)  &
$ 7.759 \pm 0.041$ & $ 7.729 \pm 0.022$ &
$ 7.276 \pm 0.064$ & $ 7.570 \pm 0.063$ \\
  \\
  Ne$^{++}$/H$^+$ ($\times$10$^5$) &
$ 0.800 \pm 0.114$ & $ 0.778 \pm 0.057$ &
$ 0.086 \pm 0.025$ & $ 0.324 \pm 0.081$ \\
  ICF &
  1.155 &  1.106 &
  1.390 &  1.309 \\
  log(Ne/O) &
$-0.793 \pm 0.098$ & $-0.795 \pm 0.047$ &
$-1.196 \pm 0.354$ & $-0.943 \pm 0.238$ \\ \tableline
  &\multicolumn{4}{c}{\sc Galaxy}\\      \cline{2-5}
  & \multicolumn{1}{c}{J0911+3135        } & \multicolumn{1}{c}{J0940+2935        } &
\multicolumn{1}{c}{J0946+5452        } & \multicolumn{1}{c}{J0956+2849 No.1     }
\\ \tableline
  $T_{\rm e}$(O {\sc iii}) (K) &
$16641  \pm2792 $ & $14889  \pm\,~923 $ &
$16192  \pm1722 $ & $19676  \pm\,~685 $  \\

  $T_{\rm e}$(O {\sc ii}) (K) &
$15200  \pm2315 $ & $14196  \pm\,~806 $ &
$14971  \pm1449 $ & $16228  \pm\,~502 $  \\

  \\
  O$^+$/H$^+$ ($\times$10$^4$) &
$ 0.243 \pm 0.096$ & $ 0.200 \pm 0.032$ &
$ 0.190 \pm 0.048$ & $ 0.028 \pm 0.002$ \\
  O$^{++}$/H$^+$ ($\times$10$^4$) &
$ 0.081 \pm 0.034$ & $ 0.250 \pm 0.041$ &
$ 0.326 \pm 0.087$ & $ 0.108 \pm 0.009$ \\
  O$^{+++}$/H$^+$ ($\times$10$^6$) &
   \nodata~~~~~    &    \nodata~~~~~    &
   \nodata~~~~~    & $ 0.352 \pm 0.047$ \\
  O/H ($\times$10$^4$) &
$ 0.325 \pm 0.101$ & $ 0.451 \pm 0.051$ &
$ 0.516 \pm 0.099$ & $ 0.139 \pm 0.009$ \\
  12 + log(O/H)  &
$ 7.511 \pm 0.136$ & $ 7.654 \pm 0.049$ &
$ 7.713 \pm 0.083$ & $ 7.144 \pm 0.028$ \\
  \\
  Ne$^{++}$/H$^+$ ($\times$10$^5$) &
$ 0.248 \pm 0.132$ & $ 0.430 \pm 0.077$ &
$ 0.674 \pm 0.187$ & $ 0.168 \pm 0.014$ \\
  ICF &
  1.398 &  1.180 &
  1.139 &  1.075 \\
  log(Ne/O) &
$-0.972 \pm 0.674$ & $-0.948 \pm 0.128$ &
$-0.827 \pm 0.187$ & $-0.888 \pm 0.053$ \\ \tableline
  &\multicolumn{4}{c}{\sc Galaxy}\\      \cline{2-5}
  &\multicolumn{1}{c}{J0956+2849 No.2     } & \multicolumn{1}{c}{J0956+2849 No.6     } &
\multicolumn{1}{c}{J2238+1400 No.1     } & \multicolumn{1}{c}{J2238+1400 No.2     }
\\ \tableline
  $T_{\rm e}$(O {\sc iii}) (K) &
$19913  \pm2184 $ & $22170  \pm4668 $ &
$20975  \pm\,~268 $ & $18580  \pm\,~260 $  \\

  $T_{\rm e}$(O {\sc ii}) (K) &
$16270  \pm1583 $ & $16396  \pm2938 $ &
$16392  \pm\,~183 $ & $15961  \pm\,~201 $  \\

  \\
  O$^+$/H$^+$ ($\times$10$^4$) &
$ 0.049 \pm 0.012$ & $ 0.096 \pm 0.042$ &
$ 0.021 \pm 0.001$ & $ 0.043 \pm 0.002$ \\
  O$^{++}$/H$^+$ ($\times$10$^4$) &
$ 0.076 \pm 0.019$ & $ 0.067 \pm 0.030$ &
$ 0.257 \pm 0.008$ & $ 0.318 \pm 0.011$ \\
  O$^{+++}$/H$^+$ ($\times$10$^6$) &
$ 0.878 \pm 0.192$ &    \nodata~~~~~    &
$ 0.295 \pm 0.024$ & $ 0.395 \pm 0.049$ \\
  O/H ($\times$10$^4$) &
$ 0.134 \pm 0.022$ & $ 0.163 \pm 0.052$ &
$ 0.282 \pm 0.008$ & $ 0.365 \pm 0.011$ \\
  12 + log(O/H)  &
$ 7.127 \pm 0.072$ & $ 7.212 \pm 0.137$ &
$ 7.450 \pm 0.012$ & $ 7.562 \pm 0.013$ \\
  \\
  Ne$^{++}$/H$^+$ ($\times$10$^5$) &
$ 0.132 \pm 0.036$ & $ 0.094 \pm 0.050$ &
$ 0.478 \pm 0.015$ & $ 0.614 \pm 0.022$ \\
  ICF &
  1.173 &  1.275 &
  1.030 &  1.042 \\
  log(Ne/O) &
$-0.936 \pm 0.189$ & $-1.132 \pm 0.464$ &
$-0.758 \pm 0.019$ & $-0.757 \pm 0.022$ \\
  \\
  Fe$^{++}$/H$^+$($\times$10$^6$)(4658) &
   \nodata~~~~~    &    \nodata~~~~~    &
$ 0.052 \pm 0.008$ &    \nodata~~~~~    \\
  Fe$^{++}$/H$^+$($\times$10$^6$)(4988) &
   \nodata~~~~~    &    \nodata~~~~~    &
$ 0.088 \pm 0.011$ &    \nodata~~~~~    \\
  ICF &
  \nodata~~~~~ &  \nodata~~~~~ &
 18.255 &  \nodata~~~~~ \\
  log(Fe/O) (4658) &
   \nodata~~~~~    &    \nodata~~~~~    &
$-1.475 \pm 0.068$ &    \nodata~~~~~    \\
  log(Fe/O) (4988) &
   \nodata~~~~~    &    \nodata~~~~~    &
$-1.244 \pm 0.055$ &    \nodata~~~~~    \\ 
  \enddata
  \end{deluxetable}

\clearpage

\begin{deluxetable}{lccccc}
\tablenum{5}
\tablecolumns{6}
\tablewidth{0pt}
\tablecaption{Oxygen abundances derived by different methods \label{tab5}}
\tablehead{
\colhead{SDSS Name}&\colhead{12+logO/H\tablenotemark{a}}
&\colhead{12+logO/H\tablenotemark{b}}&\colhead{12+logO/H\tablenotemark{c}}&\colhead{12+logO/H\tablenotemark{d}}
&\colhead{12+logO/H\tablenotemark{e}}
}
\startdata
SDSSJ0113$+$0052 No.1&7.24$\pm$0.05&7.35&7.16&7.25&7.29$\pm$0.04 \\
SDSSJ0113$+$0052 No.2&\nodata      &7.48&7.16&7.25&7.34$\pm$0.06 \\
SDSSJ0113$+$0052 No.4&\nodata      &7.44&7.20&7.30&7.35$\pm$0.05 \\
SDSSJ0204$-$1009 No.1&\nodata      &7.52&7.19&7.29&7.37$\pm$0.06 \\
SDSSJ0204$-$1009 No.2&7.56$\pm$0.03&7.60&7.38&7.47&7.55$\pm$0.05 \\
SDSSJ0254$+$0035     &\nodata      &7.45&7.11&7.20&7.28$\pm$0.05 \\
SDSSJ0301$-$0052     &7.52$\pm$0.06&7.44&7.44&7.52&7.59$\pm$0.06 \\
SDSSJ0313$+$0010     &\nodata      &7.58&7.25&7.35&7.44$\pm$0.06 \\
SDSSJ0747$+$5111 No.1&7.75$\pm$0.04&7.96&7.70&7.68&7.84$\pm$0.06 \\
SDSSJ0747$+$5111 No.2&7.73$\pm$0.02&7.78&7.62&7.64&7.78$\pm$0.06 \\
SDSSJ0812$+$4836     &\nodata      &7.35&7.08&7.16&7.28$\pm$0.06 \\
SDSSJ0859$+$3923     &\nodata      &7.69&7.36&7.45&7.57$\pm$0.06 \\
SDSSJ0911$+$3135     &7.51$\pm$0.14&7.46&7.24&7.34&7.46$\pm$0.05 \\
SDSSJ0940$+$2935     &7.65$\pm$0.05&7.65&7.35&7.45&7.53$\pm$0.04 \\
SDSSJ0946$+$5452     &7.71$\pm$0.08&7.88&7.64&7.66&7.80$\pm$0.06 \\
SDSSJ0956$+$2849 No.1&7.14$\pm$0.03&7.16&7.04&7.12&7.13$\pm$0.04 \\
SDSSJ0956$+$2849 No.2&7.13$\pm$0.07&7.26&7.00&7.03&7.10$\pm$0.03 \\
SDSSJ0956$+$2849 No.6&7.21$\pm$0.14&7.45&7.12&7.21&7.28$\pm$0.04 \\
SDSSJ2238$+$1400 No.1&7.45$\pm$0.01&7.51&7.66&7.66&7.78$\pm$0.07 \\
SDSSJ2238$+$1400 No.2&7.56$\pm$0.01&7.57&7.66&7.66&7.79$\pm$0.07 \\ \tableline
$\Delta$(12+logO/H)\tablenotemark{f}& \nodata     &0.10&0.13&0.07&0.06 \\
\enddata
\tablenotetext{a}{$T_e$(O {\sc iii}) is derived from the 
[O {\sc iii}] $\lambda$4363/($\lambda$4959+$\lambda$5007) ratio 
(direct method).}
\tablenotetext{b}{12 + logO/H is derived from Eq. \ref{eq:o1} \citep{PT05}
(empirical method).}
\tablenotetext{c}{12 + logO/H is derived from Eq. \ref{eq:o2} \citep{N06}
(empirical method).}
\tablenotetext{d}{12 + logO/H is derived from Eq. \ref{eq:o3} \citep{Y07}
(empirical method).}
\tablenotetext{e}{$T_e$(O {\sc iii}) is derived from Eq. \ref{eq:toiii}
(semi-empirical method).}
\tablenotetext{f}{Average difference between the empirical abundances and 
those derived by the direct method, excluding the two H {\sc ii} regions in J2238+1400.}
\end{deluxetable}

\clearpage

\clearpage

\begin{figure}
\figurenum{2}
\plotfiddle{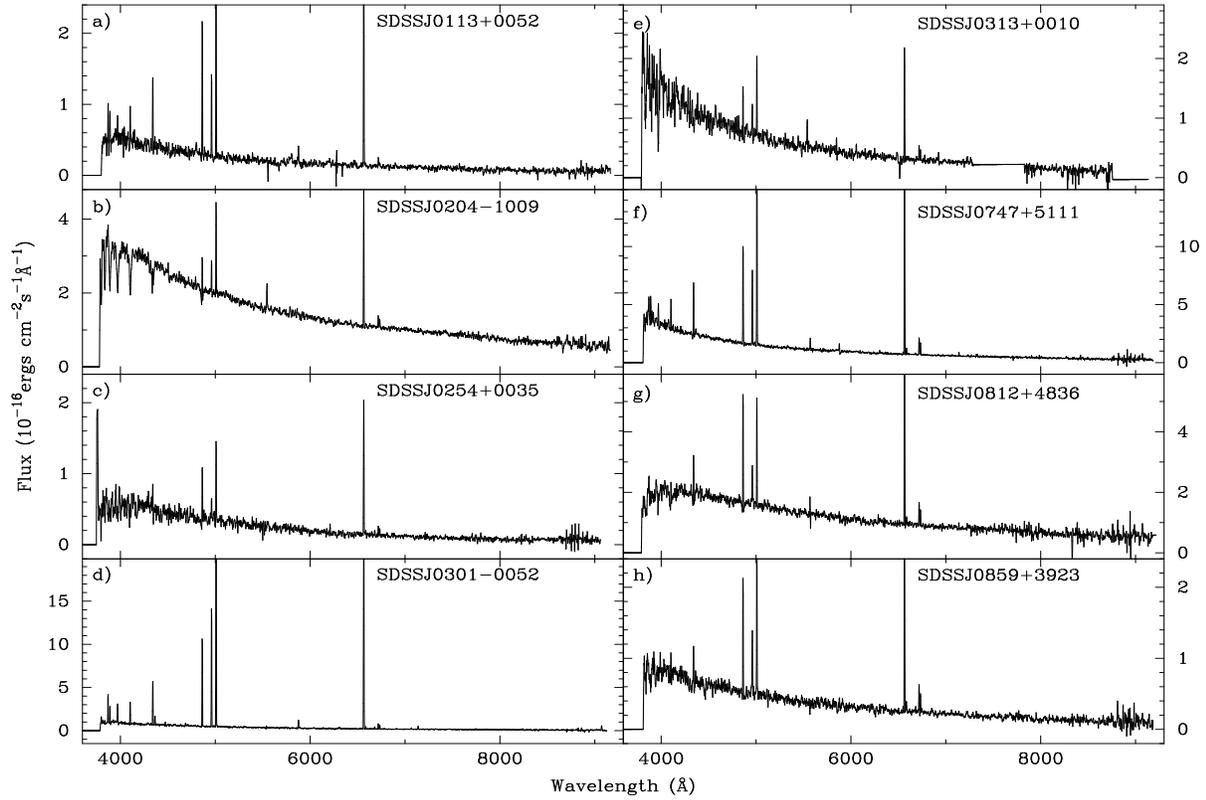}{1.pt}{-90.}{300.}{450.}{0.}{0.}
\caption{SDSS spectra of galaxies in MMT subsample. }
\label{fig2}
\end{figure}

\clearpage

\begin{figure}
\figurenum{2}
\plotfiddle{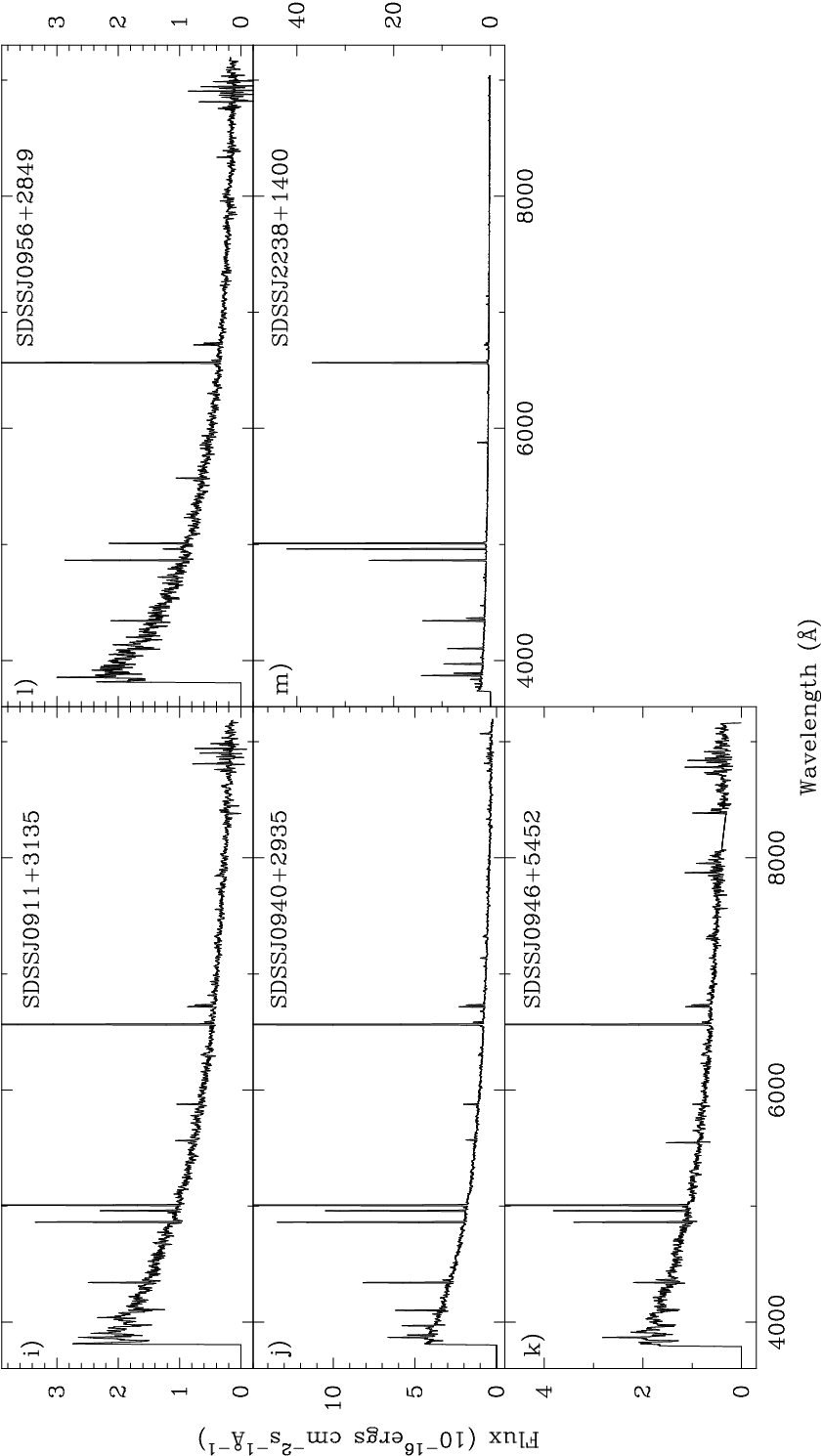}{1.pt}{-90.}{250.}{450.}{0.}{0.}
\caption{Continued.}
\end{figure}

\clearpage

\begin{figure}
\figurenum{3}
\plotfiddle{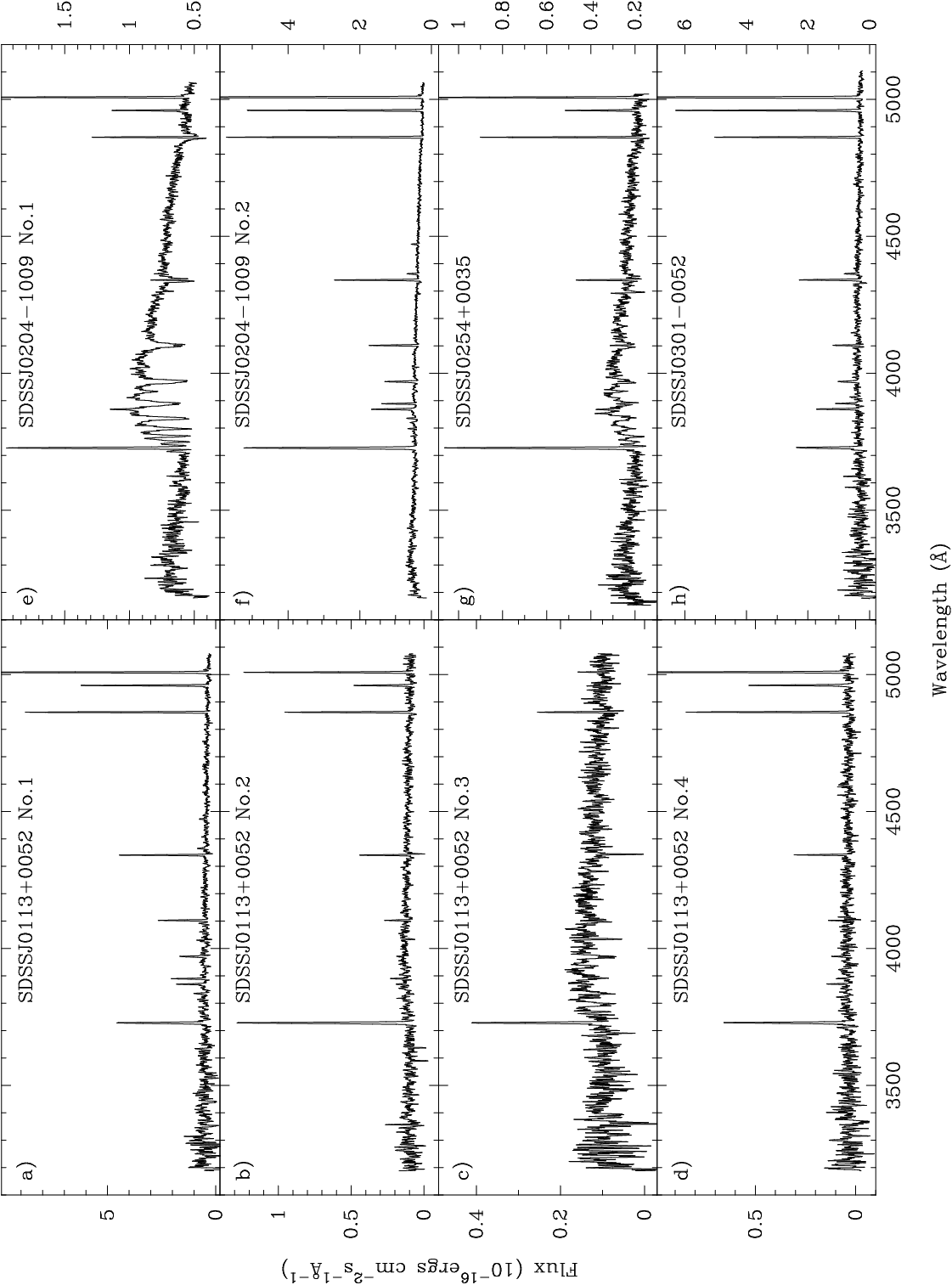}{1.pt}{-90.}{300.}{450.}{0.}{0.}
\caption{MMT spectra of extremely low-metallicity SDSS galaxies. The inset
in panel t shows a blow-up of the blue part of the spectrum.}
\label{fig3}
\end{figure}

\clearpage

\begin{figure}
\figurenum{3}
\plotfiddle{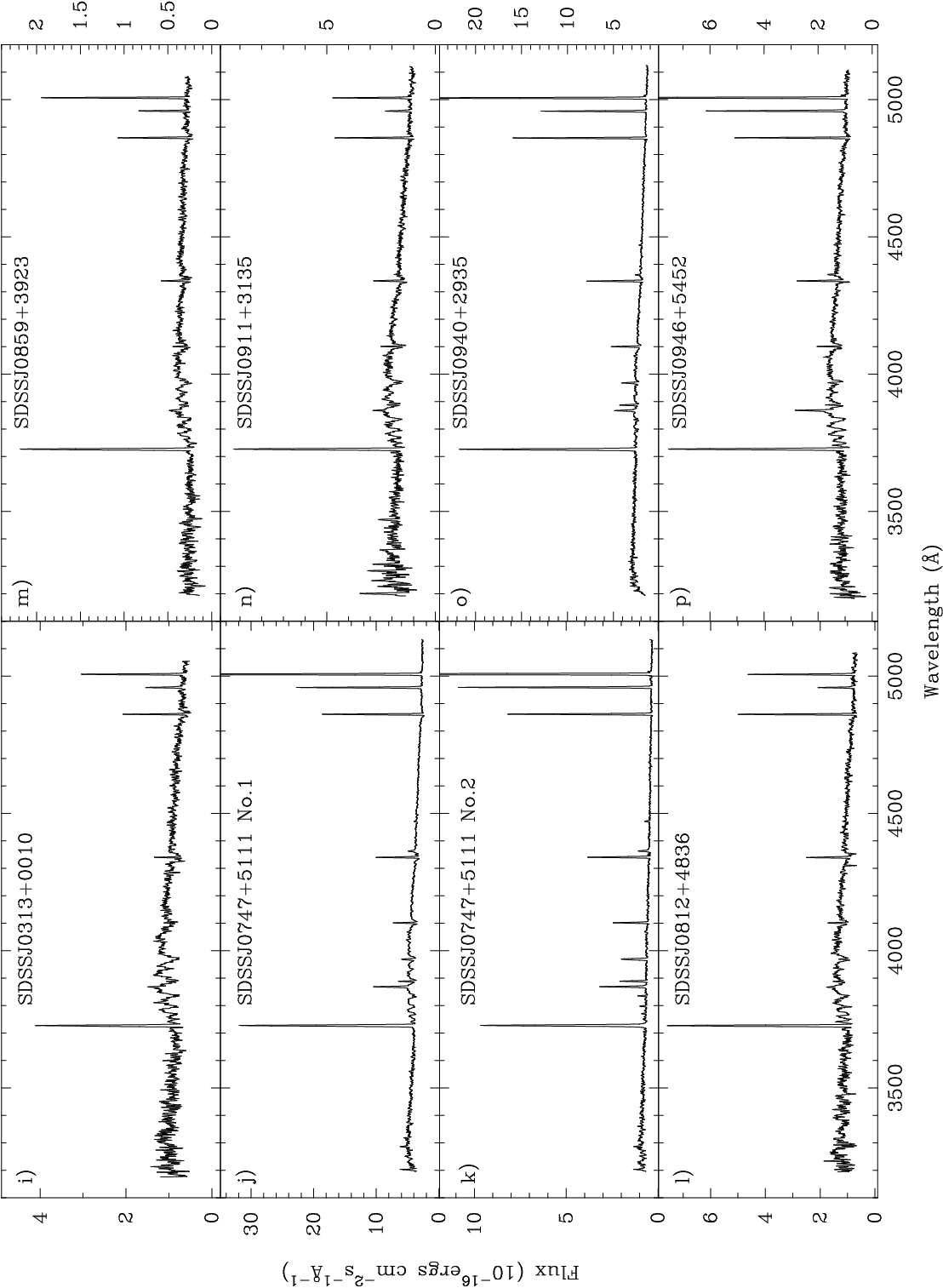}{1.pt}{-90.}{300.}{450.}{0.}{0.}
\caption{Continued.}
\end{figure}

\clearpage

\begin{figure}
\figurenum{3}
\plotfiddle{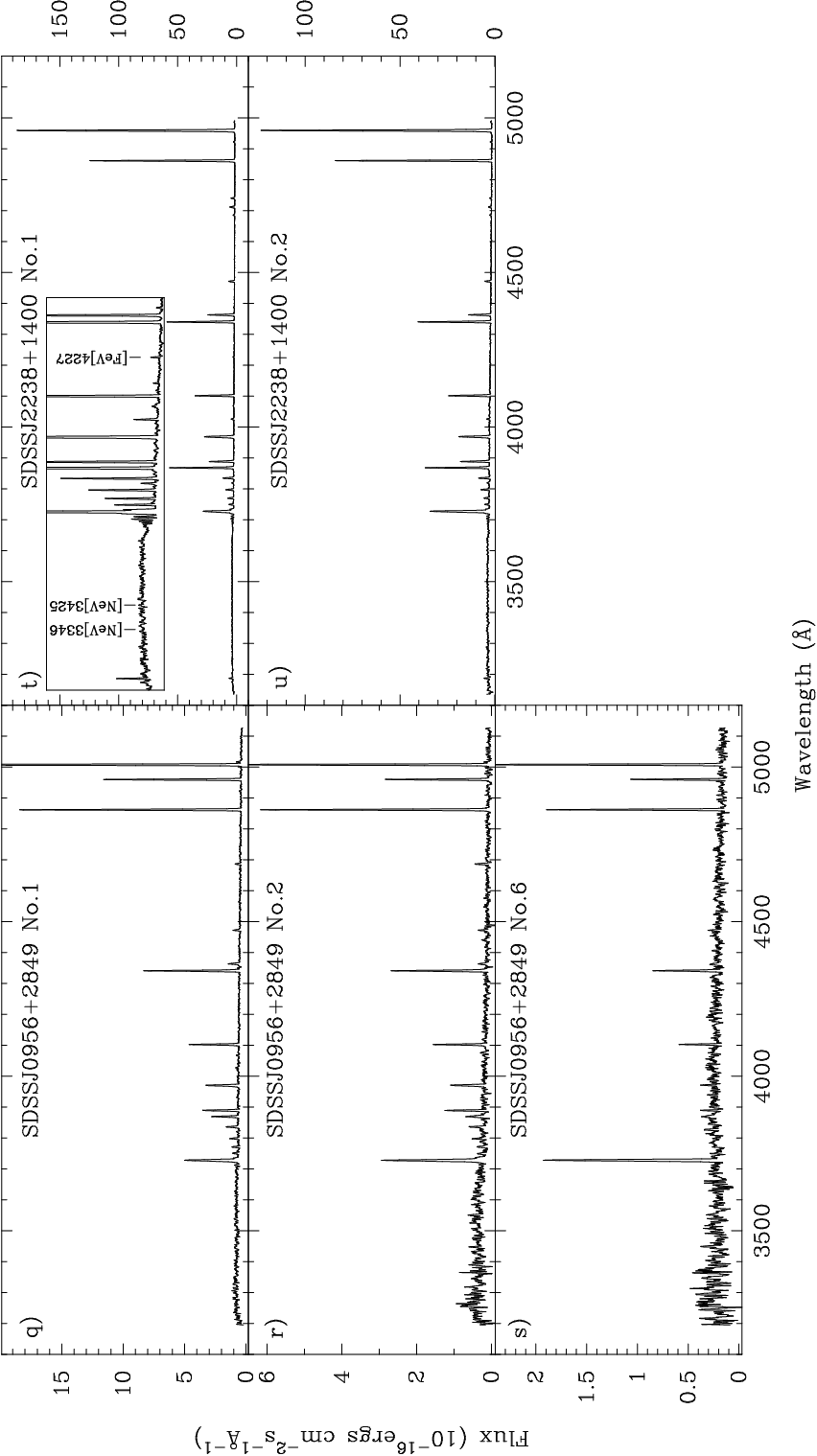}{1.pt}{-90.}{250.}{450.}{0.}{0.}
\caption{Continued.}
\end{figure}

\end{document}